\documentclass[aps,prl,twocolumn,footinbib,floatfix,superscriptaddress]{revtex4-2}

\usepackage{upgreek}

\usepackage{graphicx}
\usepackage{indentfirst}
\usepackage{braket}
\usepackage{float}
\usepackage{amsmath}
\usepackage{bm}
\usepackage{physics}
\usepackage{amssymb}
\usepackage{CJK}
\usepackage{esint}
\usepackage{color}
\usepackage[T1]{fontenc}
\usepackage{subfigure}
\usepackage{amsfonts}
\usepackage{footmisc}
\usepackage{scrextend}
\usepackage{multirow}
\usepackage[outdir=./]{epstopdf}
\usepackage{svg}

\usepackage{float}
\makeatletter
\let\newfloat\newfloat@ltx
\makeatother
\usepackage{algorithm}
\usepackage{algpseudocode}

\usepackage[english]{babel}
\usepackage{url}
\usepackage{comment}
\usepackage{array}
\usepackage{subfiles}
\usepackage{tikz}
\usetikzlibrary{shapes,arrows}
\usepackage{mathrsfs}
\usepackage[mathscr]{eucal}
\usepackage[hyperfootnotes=false]{hyperref}
\usepackage{cleveref}
\usepackage{stmaryrd}
\definecolor{darkblue}{rgb}{0,0,1/2}
\hypersetup{
colorlinks=true,
linkcolor=black,
filecolor=blue,
citecolor=darkblue,  
urlcolor=black,
}

\newtheorem{theorem}{Theorem}

\newcommand{\calA}{{\cal A}}

\newcommand{\calL}{{\cal L}}

\newcommand{\1}{^{(1)}}

\def\be{\begin{equation}}
\def\ee{\end{equation}}
\def\ba{\begin{eqnarray}}
\def\ea{\end{eqnarray}}
\def\bal{\begin{equation}\begin{aligned}}
\def\eal{\end{aligned}\end{equation}}

\usepackage[normalem]{ulem}

\begin{document}



\title{Quantum transduction via generalized continuous-variable teleportation}
\begin{abstract}
Quantum transduction converts quantum states between different frequencies. Similarly, quantum teleportation transfers quantum states between different systems. While often appreciated for quantum communication between distant locations, teleportation can also operate between different frequencies. In terms of bosonic modes, quantum teleportation relies on two-mode squeezing---squeezed on the balanced Einstein-Podolsky-Rosen (EPR) quadratures---for entanglement and a balanced beamsplitter plus homodyne detection for Bell measurement. Since cross-frequency band entanglement is challenging to generate,
we propose to enable transduction with teleportation, while only relying on a conventional transduction device and narrow-band entanglement. Our insight is that a transduction device performs a cross-frequency-band beamsplitter, almost what is needed in a Bell measurement, except that it is often not balanced. We resolve the issue by proposing a generalized teleportation protocol with arbitrarily unbalanced EPR quadratures and matched unbalanced beamsplitters. The protocol can enhance quantum transduction, using only off-the-shelf components---intraband beamsplitters, offline squeezing, homodyne detection and displacements. The proposed protocol achieves perfect transduction when applying to a transduction device with an arbitrarily low efficiency, at infinite squeezing.  A minimum squeezing in the range of $[1.195,4.343]$ decibels is needed to achieve quantum capacity enhancement.
\end{abstract}

\author{Quntao Zhuang}
\email{qzhuang@usc.edu}

\address{
Ming Hsieh Department of Electrical and Computer Engineering, University of Southern California, Los
Angeles, California 90089, USA
}
\address{
Department of Physics and Astronomy, University of Southern California, Los
Angeles, California 90089, USA
}

\maketitle

Quantum transduction aims to interconnect quantum computers and processors via transferring quantum states between different frequencies~\cite{lauk2020perspectives,awschalom2021development,han2021microwave}. In particular, microwave-optical quantum transduction connects quantum devices that operate under microwave control, e.g. microwave superconducting qubits and semiconductor spin qubits, towards the optical telecommunication photons.  
Despite proposals based on various physical platforms~\cite{andrews2014bidirectional,bochmann2013nanomechanical,vainsencher2016bi,balram2016coherent,Tsang2010,Tsang2011,fan2018superconducting,xu2020bidirectional,jiang2020efficient,PhysRevLett.103.043603,PhysRevLett.113.203601,shao2019microwave,fiaschi2021optomechanical,han2020cavity,zhong2020proposal,mirhosseini2020superconducting,forsch2020microwave}, current transduction systems are still far from satisfactory, hampered by a conundrum to balance transduction efficiency, pump-induced heating, and bandwidth~\cite{holzgrafe2020cavity,mirhosseini2020superconducting,sahu2022quantum,brubaker2022optomechanical,qiu2023coherent,sahu2023entangling}. 

Quantum teleportation~\cite{bennett1993} seems to be a perfect solution to transduction, as it transfers a quantum state from a sender to a receiver, without sending a physical qubit through. To enable the transfer of bosonic modes, continuous-variable (CV) quantum teleportation~\cite{braunstein1998,pirandola2006,furusawa1998unconditional} adopts the CV analog of Bell state---the two-mode squeezed vacuum---as the entanglement source and balanced beamsplitter plus homodyne as the Bell measurement. 
However, teleportation relies on preshared quantum entanglement, and generating entanglement across different frequency bands (e.g., microwave-optical entanglement) is almost as hard as transduction itself~\cite{sahu2023entangling}, making the previous proposal relying on microwave-optical entanglement~\cite{wu2021deterministic} challenging. 





In this work, we apply teleportation to enable quantum transduction, without requiring cross-frequency-band entanglement.
Our insight is that a conventional transduction device performs a cross-frequency-band beamsplitter, almost what is needed in a Bell measurement, except that it is often not balanced. We resolve the issue by proposing a generalized CV teleportation protocol that removes the restriction on balanced mixing in the EPR quadratures and the Bell measurement. The proposed generalized-CV-teleportation-based quantum transduction (CV-tele-QT) protocol increases transduction efficiency with minimum added noise, while only relying on offline single-mode squeezing, intraband operations and feedforward control. In the limit of infinite squeezing, it allows boosting the transduction efficiency to unity from an arbitrary small value. When applied to microwave-to-optical transduction, only offline optical squeezing is needed; when applied to optical-to-microwave transduction, only offline microwave squeezing is needed. In both cases, a minimum squeezing in the range of $[1.195,4.343]$ decibels is needed to achieve transduction quantum capacity enhancement.

{\em Preliminary.---} A mode of a CV quantum systems involves the annihilation and creation operators, $\hat{a}$ and $\hat{a}^\dagger$, which satisfy the canonical communication relation $[\hat{a},\hat{a}^\dagger]=1$. Correspondingly, the position and momentum quadrature operators 
$ 
\hat{q}=\hat{a}+\hat{a}^\dagger, \hat{p}=(\hat{a}-\hat{a}^\dagger)/i,
$ 
satisfy the commutation relation $[\hat{q},\hat{p}]=2i$, where we have let $\hbar=2$. Under such a choice of units, the vacuum fluctuation ${\rm var}(\hat{q})={\rm var}(\hat{p})=1$. Common quantum channels (Gaussian channels~\cite{Weedbrook2012}) can be described by linear transforms of annihilation operators. In particular, a bosonic pure-loss channel $\calL_\eta$ with transmissivity $\eta$ maps $\hat{a}\to \sqrt{\eta}\hat{a}+\sqrt{1-\eta}\hat{v}$, where $\hat{v}$ is vacuum. A general Gaussian additive noise channel $\calA_{\sigma_q^2,\sigma_p^2}$ maps $\hat{a}\to \hat{a}+\xi$, where ${\rm Re}(\xi)$ and ${\rm Im}(\xi)$ are zero-mean Gaussian random variables with variances $\sigma_q^2/4,\sigma_p^2/4$. When the noise is symmetric, $\sigma_q^2=\sigma_p^2=\sigma^2$, we can simplify the notation as $\calA_{\sigma^2}\equiv \calA_{\sigma^2,\sigma^2}$.

\begin{figure}[t]
    \centering
    \includegraphics[width=\linewidth]{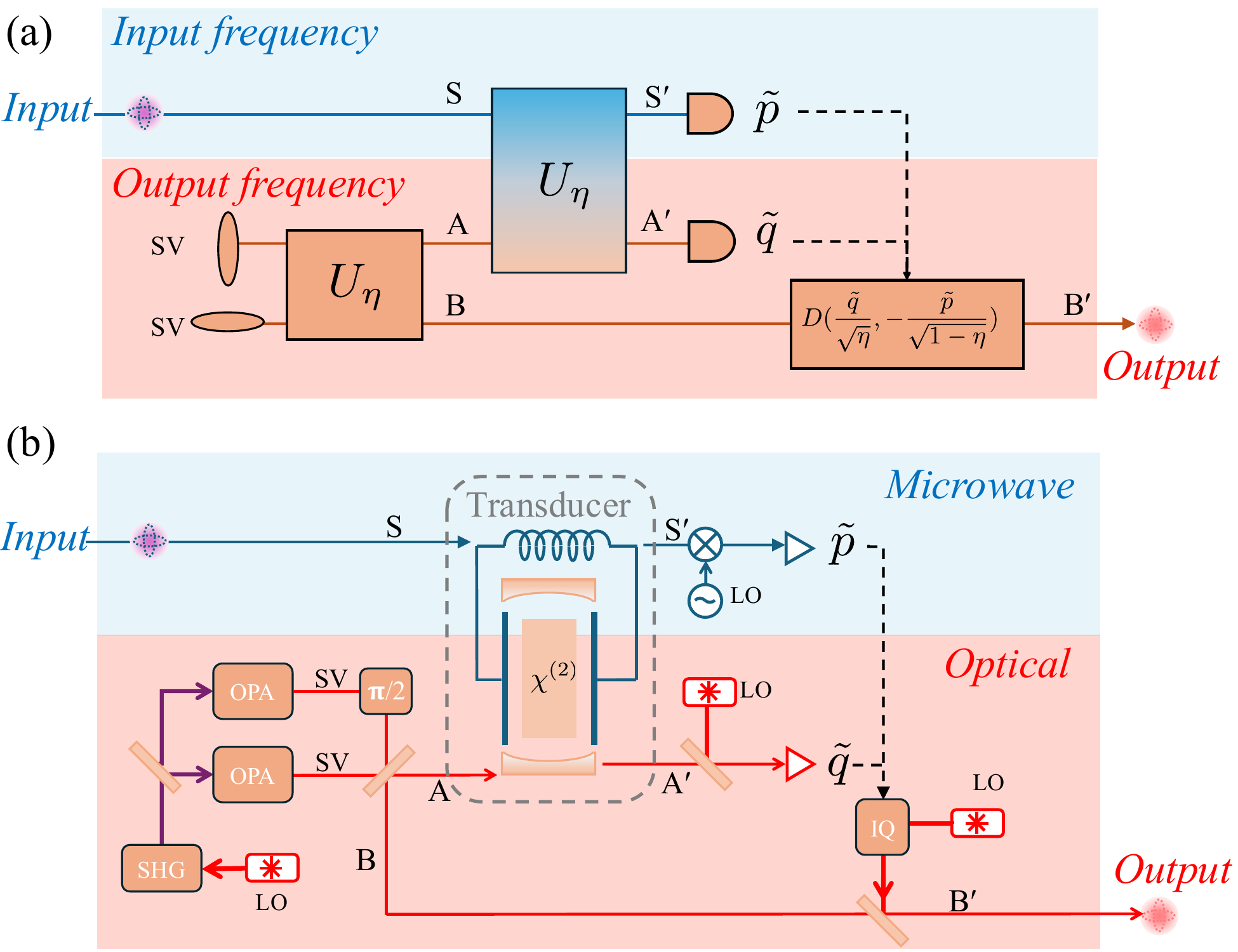}
    \caption{(a) The generalized continuous-variable quantum teleportation protocol. $U_\eta$ denotes the beamsplitter with ratio $\eta$. (b) Example of CV-tele-QT applied to microwave-to-optical transduction. LO: local oscillator. SHG: second harmonic generation. OPA: optical parametric amplifier. SV: single-mode squeezing. IQ: optical modulator. The box with $\pi/2$ represents a phase shift of $\pi/2$.}
    \label{fig:schematic}
\end{figure}

{\em Generalized CV teleportation.---} As shown in Fig.~\ref{fig:schematic}(a), the protocol starts with preparing the entangled state $\ket{\psi}_{AB}$ via mixing two single-mode squeezed vacuums, squeezed under orthogonal directions, under a beamsplitter with splitting ratio $\eta$. This results in suppressed variances ${\rm var}(\hat{q}_{-,\eta})={\rm var}(\hat{p}_{+,\eta})=1/G$, where the generalized EPR quadratures 
\begin{subequations}
\begin{align} 
&\hat{q}_{-,\eta}\equiv \sqrt{1-\eta}\hat{q}_A-\sqrt{\eta}\hat{q}_B,
\\
&\hat{p}_{+,\eta}\equiv \sqrt{\eta}\hat{p}_A+\sqrt{1-\eta}\hat{p}_B,
\end{align}
\label{generalized_EPR}
\end{subequations}
and $G$ is the squeezing gain. In the case of a balanced beamsplitter $\eta=1/2$, $\ket{\psi}_{AB}$ is a conventional two-mode squeezed vacuum. In the general case, $\ket{\psi}_{AB}$ is a zero-mean two-mode entangled Gaussian state~\cite{Weedbrook2012}.

With the entanglement distributed, the sender then picks mode $A$ from the entangled pair and mixes it with the signal $S$ on a beamsplitter with ratio $\eta$, which transforms
\begin{subequations}
\begin{align}
&\hat{a}_{S'}=\sqrt{\eta}\hat{a}_A+\sqrt{1-\eta}\hat{a}_S,
\\
&\hat{a}_{A'}=-\sqrt{\eta}\hat{a}_S+\sqrt{1-\eta} \hat{a}_A.
\end{align}
\label{AS_beamsplitter}
\end{subequations}
Finally, the sender performs homodyne measurement on $S^\prime$ and $A^\prime$, leading to measurement results $\tilde{p}$ on momentum quadrature of $S^\prime$ and $\tilde{q}$ on position quadrature of $A^\prime$. The sender then sends the classical measurement results to the receiver, who performs a displacement on the mode $B$, with quadrature mean shifts $\tilde{q}/\sqrt{\eta}$ and $-\tilde{p}/\sqrt{1-\eta}$ on position and momentum quadratures, creating the quantum state for the output $B^\prime$.

As we detail in the appendix, we prove the following result:
\begin{theorem}
\label{theorem:cv_teleportation}
(Generalized CV teleportation) In the infinite squeezing limit of the generalized CV teleportation protocol, the output state equals the unknown input state,
\be 
\hat{\rho}_{B^\prime}=\hat{\rho}_S, G\to\infty.
\label{equality_ginf}
\ee 
In the case of finite squeezing, the final output state equals the input state up to additive Gaussian noises, with variances $1/(\eta G)$ on the position quadrature and $1/[(1-\eta)G]$ on the momentum quadrature:
\be 
\hat{\rho}_{B^\prime}=\calA_{1/(\eta G),1/[(1-\eta)G]}(\hat{\rho}_S).
\label{add_channel_G}
\ee 
\end{theorem}



{\em Enhancing quantum transduction.---}Having established the generalized teleportation scheme, we now explore its potential to address a critical challenge in quantum technology---efficient quantum transduction. As shown in Fig.~\ref{fig:schematic}(a), quantum transduction aims to convert the input quantum state at an input frequency (indicated by the blue color background) to the output at an output frequency (indicated by the red color background). A direct quantum transduction device (indicated by the blue-red blend box) essentially implements a beamsplitter between the input and output frequencies (Eq.~\eqref{AS_beamsplitter}), producing the output mode ${A'}$ from the input mode $S$, with transduction efficiency $\eta$. Fig.~\ref{fig:schematic}(b) dashed box depicts an electro-optics quantum transducer as an example~\cite{fan2018superconducting}, where an optical cavity with $\chi^{(2)}$-nonlinear material is placed between capacitors of a LC microwave resonator. While the direct transduction approach takes $A$ as vacuum, we can engineer $A$ to be entangled with another ancilla mode $B$, as prescribed by Eq.~\eqref{generalized_EPR}. Note that both $A$ and $B$ are in the output frequency band---they can be microwave-microwave entanglement or optical-optical entanglement. Now one performs the homodyne measurements on $S^\prime$ at the input frequency and on $A^\prime$ at the output frequency. Finally, the CV-tele-QT concludes with the corrective displacement at the output frequency.

The proposed CV-tele-QT protocol only requires off-the-shelf components: offline squeezing and beamsplitter at a single frequency band, homodyne detection, displacement and a conventional quantum transduction device. Therefore, both microwave-to-optical and optical-to-microwave direction can be realized under current experimental capabilities. Fig.~\ref{fig:schematic}(b) provides a design of the experimental system for microwave-to-optical CV-tele-QT, where we choose electro-optical transduction device as an example. At the same time, the protocol does not require any specific quantum information encoding.
Theorem~\ref{theorem:cv_teleportation} immediately implies that with an infinite amount of microwave-microwave or optical-optical squeezing, one can achieve perfect quantum transduction via CV-tele-QT. Below, we evaluate the performance at finite squeezing.

\begin{figure}[t]
    \centering
    \includegraphics[width=\linewidth]{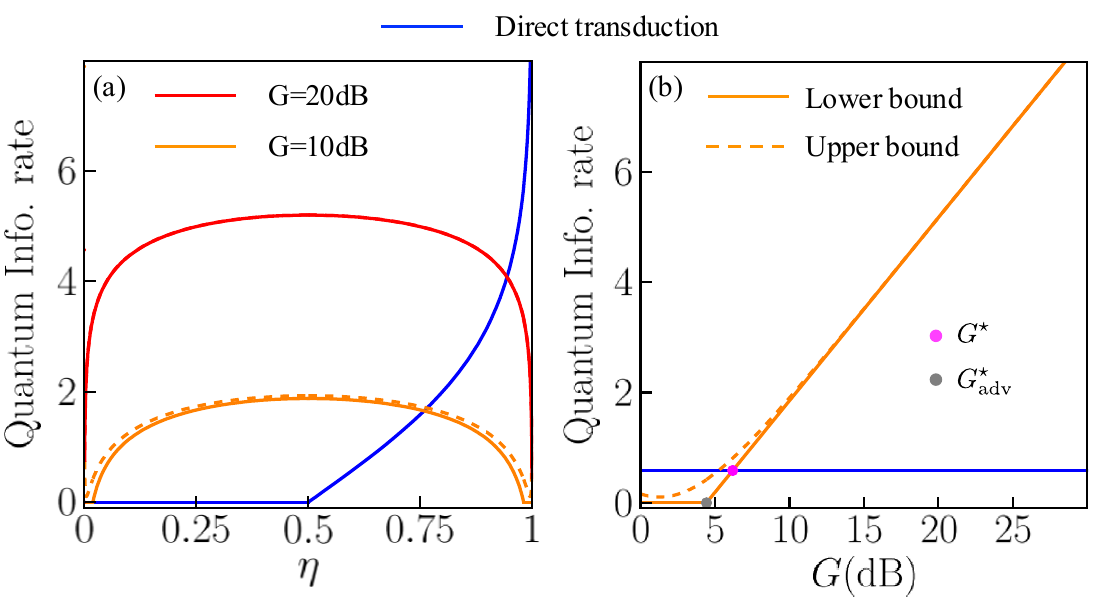}
    \caption{(a) Quantum information rate versus the direct transduction efficiency $\eta$ for direct transduction (blue solid), CV-tele-QT with $G=10$dB (orange) and with $G=20$dB (red). (b) Quantum information rate of CV-tele-QT versus the squeezing gain $G$ in dB at $\eta=0.6$. Lower bounds are plot in solid and up bounds dashed. $G^\star$ and $G^\star_{\rm adv}$ are threshold gain values in Eqs.~\eqref{G_star} and \eqref{G_star_adv}.}
    \label{fig:rate_comparison}
\end{figure}

{\em Finite squeezing analyses of CV-tele-QT.---}
At finite squeezing, the quantum channel of CV-tele-QT is an additive noise channel $\calA_{1/(\eta G),1/[(1-\eta)G]}$ in Eq.~\eqref{add_channel_G}, with asymmetric noise variances. To compare with direct transduction, where a bosonic pure-loss channel $\calL_\eta$ with efficiency $\eta$ is implemented, we evaluate the quantum capacity~\cite{lloyd1997capacity,shor2002quantum,devetak2005private}.
The quantum capacity of a bosonic pure-loss channel can be exactly solved as~\cite{holevo2001evaluating}
\be 
Q\left(\calL_\eta\right)=\max\left[\log_2(\frac{\eta}{1-\eta}),0\right].
\ee 
In terms of quantum capacity without energy constraint, $Q\left(\calA_{1/(\eta G),1/[(1-\eta)G]}\right)=Q\left(\calA_{1/[\sqrt{\eta(1-\eta)} G]}\right)$ equals the capacity of a Gaussian additive noise channel with symmetric variance as one can introduce single-mode squeezing as pre- and post-processing~\cite{wu2021continuous}.
Then, we can apply exisiting upper~\cite{fanizza2021estimating} and lower bounds, $Q_{\rm LB}\left(\calA_v\right)\le Q\left(\calA_v\right) \le Q_{\rm UB}\left(\calA_v\right)$, where
\small
\begin{align}
&Q_{\rm LB}\left(\calA_v\right)= \max\left[-\log_2\left(\frac{v}{2}\right)-\frac{1}{\ln(2)} ,0\right],
\label{Q_LB}
\\
&Q_{\rm UB}\left(\calA_v\right)=\max\left[-\log_2\left(\frac{v}{2}\right)-\frac{1}{\ln(2)}+2h\left(\sqrt{1+\frac{v^2}{4}}\right),0\right],
\label{Q_UB}
\end{align}
\normalsize
where the quadrature variance $v\ge1$ and $h(x):=[(x+1)/2]\log_2[(x+1)/2]-[(x-1)/2]\log_2[(x-1)/2]$.

In Fig.~\ref{fig:rate_comparison}(a), we plot the quantum information rate of the CV-tele-QT with squeezing gain $G=10$dB (orange) and $G=20$dB (red) in comparison with the direct transduction capacity $Q\left(\calL_\eta\right)$ (blue), versus the efficiency of the transduction device $\eta$. The lower bound $Q_{\rm LB}\left(\calA_{1/[\sqrt{\eta(1-\eta)} G]}\right)$ and upper bound $Q_{\rm UB}\left(\calA_{1/[\sqrt{\eta(1-\eta)} G]}\right)$ are plotted in solid and dashed correspondingly. We identify a positive quantum information rate of CV-tele-QT for an arbitrary level of transduction-device efficiency $\eta$ for the CV-tele-QT. In comparison, the direct transduction has $Q=0$ for $\eta\le1/2$. Therefore, in the region of $\eta\le 1/2$, the CV-tele-QT enjoys drastic advantages over direct transduction. 

\begin{figure}[t]
    \centering
    \includegraphics[width=0.7\linewidth]{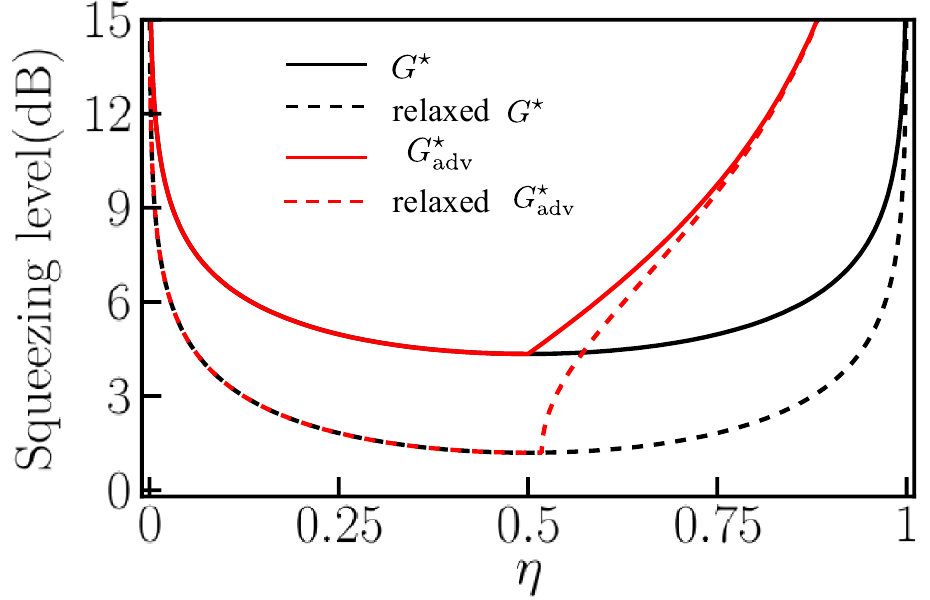}
    \caption{Squeezing thresholds in decibels (dB). Black solid is the threshold $G^\star$ to guarantee positive quantum capacity lower bound (Ineq.~\eqref{G_star}) and red solid is the threshold $G^\star_{\rm adv}$ to guarantee advantage of over direct quantum transduction (Ineq.~\eqref{G_star_adv}). The dashed curves are optimistic versions of the thresholds evaluated with the quantum capacity upper bound. The actual squeezing to enable positive quantum capacity and advantage over direct quantum transduction is in the corresponding filled regions.}
    \label{fig:squeezing_threshold}
\end{figure}

It is worth noting that both the upper and lower bounds of the CV-tele-QT are symmetric between $\eta$ and $1-\eta$, as the protocol relies on the mixing between the two modes at the transduction device, rather than a one-way conversion. Indeed the lower and upper bounds of CV-tele-QT are both maximized at $\eta=1/2$, which corresponds to the maximal interaction between the two frequency modes. 

In Fig.~\ref{fig:rate_comparison}(b), we plot the quantum information rate of CV-tele-QT versus the squeezing gain $G$, with the transduction-device efficiency $\eta=0.6$. The lower bound shows a linear increase with the number of decibels of squeezing. Indeed, the lower bound Ineq.~\eqref{Q_LB} simplifies to
\be 
Q_{\rm LB}=\max[0,\log_2({2\sqrt{\eta(1-\eta)}})+\log_2(G) -1/\ln(2)],
\ee 
showing a linear dependence on $\log_2(G)$. In particular, $Q_{\rm LB}\ge0$ as long as 
\be 
G\ge G^\star\equiv \frac{e}{2\sqrt{\eta(1-\eta)}},
\label{G_star}
\ee 
and $Q_{\rm LB}\ge Q(\calL_\eta)$ has advantage over the direct transduction when
\be 
G\ge \left\{ \begin{array}{ll}
         G^{\star}_{\rm adv}\equiv\frac{e\sqrt{\eta}}{2(1-\eta)^{3/2}} & \mbox{if $\eta \geq 1/2$};\\
        G^\star & \mbox{if $\eta\le 1/2$}.\end{array} \right. 
\label{G_star_adv}
\ee 
The threshold values $G^\star$ and $G^{\star}_{\rm adv}$ are marked in Fig.~\ref{fig:rate_comparison}(b), as gray and magenta dots. In Fig.~\ref{fig:squeezing_threshold}, we plot the thresholds~\eqref{G_star} and~\eqref{G_star_adv} as the black and red solid curves. While these are guarantees obtained from the lower bound of quantum capacity, it is possible that the actual required squeezing to have positive quantum capacity or advantage over direct transduction can be lowered. The most optimistically relaxed estimation (dashed in Fig.~\ref{fig:squeezing_threshold}) can be obtained from the quantum capacity upper bound $Q_{\rm UB}\left(\calA_{1/[\sqrt{\eta(1-\eta)} G]}\right)$. The thresholds for the actual quantum capacity lie in between the solid and dashed curves, as indicated by the color filling.  As an example, at $\eta=1/2$, the actual squeezing threshold for positive quantum capacity (which equals the threshold of advantage as $Q(\calL_\eta)=0$) is in the range of $[1.195,4.343]$ decibels.

{\em Experimental imperfections.---} To evaluate the experimental feasibility of the protocol, we consider experimental imperfections, including squeezing generation loss, transduction device loss and finite detector efficiency. In the presence of loss $1-\kappa_S$ in the squeezing generation, the generalized EPR quadratures in Eqs.~\eqref{generalized_EPR} has variances
${\rm var}(\hat{q}_{-,\eta})={\rm var}(\hat{p}_{+,\eta})=\kappa_S/G+(1-\kappa_S)$, while the corresponding anti-squeezed quadrature variances ${\rm var}(\hat{q}_{+,\eta})={\rm var}(\hat{p}_{-,\eta})=\kappa_SG+(1-\kappa_S)$. Therefore, squeezing generation loss essentially models the unbalance between squeezing and anti-squeezing in practical squeezed light source. For the loss in the transduction device and homodyne detection, we can combine them together as an overall loss $1-\kappa_H$, such that the homodyne detection is on modes $\sqrt{\kappa_H}\hat{a}_{S^\prime}+\sqrt{1-\kappa_H}\hat{a}_{V1}$ and $\sqrt{\kappa_H}\hat{a}_{A^\prime}+\sqrt{1-\kappa_H}\hat{a}_{V2}$, where $\hat{a}_{V1}$ and $\hat{a}_{V2}$ are vacuum modes. To accommodate the imperfections, we adjust the corrective displace to ${\tilde{\bm d}}_B^L=(\tilde{q}/\sqrt{\kappa_H\eta},-\tilde{p}/\sqrt{\kappa_H(1-\eta)})$. We have the following generalization of Theorem~\ref{theorem:cv_teleportation}.
\begin{theorem}
\label{theorem:cv_teleportation_imperfect}
(Imperfect generalized CV teleportation) 
In the case of finite squeezing gain $G$, homodyne efficiency $\kappa_H$ and squeezing generation efficiency $\kappa_S$, the final output state equals the input state up to additive Gaussian noises, with variances 
\begin{subequations}
\begin{align}
&\sigma_q^2=\left[\kappa_H\kappa_S/G+\left(1-\kappa_H\kappa_S\right)\right]/(\eta\kappa_H)
\\
&\sigma_p^2=\left[\kappa_H\kappa_S/G+\left(1-\kappa_H\kappa_S\right)\right]/[(1-\eta)\kappa_H].
\end{align} 
\end{subequations}
on the position and momentum quadratures correspondingly:
$
\hat{\rho}_{B^\prime}=\calA_{\sigma_q^2,\sigma_p^2}(\hat{\rho}_S).
$
\end{theorem}
It is easy to verify that when $\kappa_H=\kappa_S=1$, Theorem~\ref{theorem:cv_teleportation_imperfect} reduces back to Theorem~\ref{theorem:cv_teleportation}. 

\begin{figure}[t]
    \centering
    \includegraphics[width=\linewidth]{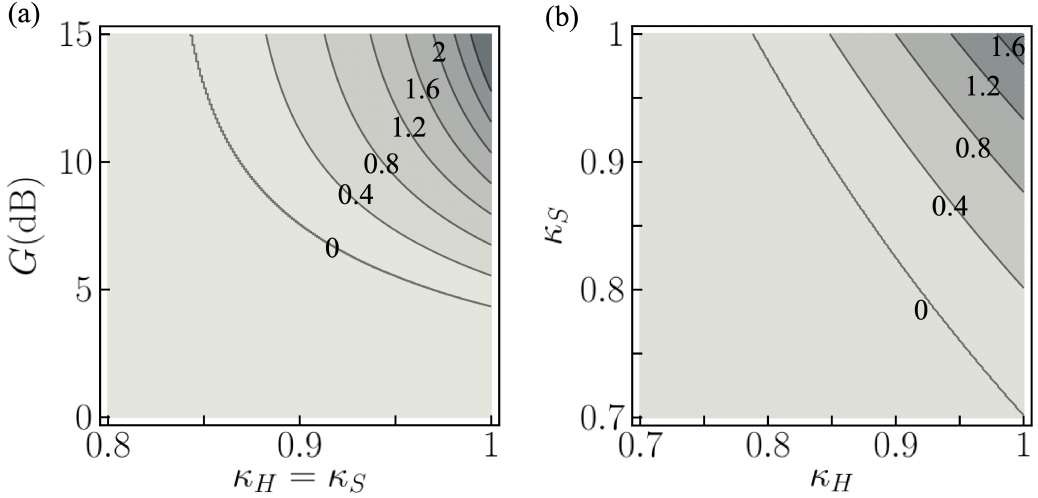}
    \caption{Quantum information rate (in bits per channel use) lower bound for CV-tele-QT for transduction efficienccy $\eta=0.5$. (a) We plot versus the squeezing gain $G$ in dB and efficiency $\kappa_H=\kappa_S$. (b) We plot versus the homodyne efficiency $\kappa_H$ and squeezing efficiency $\kappa_S$, when squeezing gain $G=10$dB. }
    \label{fig:rate_imperfection}
\end{figure}

Similar to previous analyses, the quantum capacity without energy constraint, $Q\left(\calA_{\sigma_q^2,\sigma_p^2}\right)=Q\left(\calA_{\sigma_q\sigma_p}\right)$.
Now we evaluate the quantum information rate lower bound $Q_{\rm LB}(\calA_{\sigma_p\sigma_q})$, with the noise specified in Theorem~\ref{theorem:cv_teleportation_imperfect}. To simplify the analyses, we consider $\eta=0.5$, where the direct quantum transduction has zero capacity---any positive quantum capacity lower bound indicates advantage over direct transduction. In Fig.~\ref{fig:rate_imperfection}, we see that CV-tele-QT enables a positive quantum communication rate robust to experimental imperfections. For $G=10$dB of raw squeezing, positive rate is achieved when $\kappa_H\ge 0.95$ and $\kappa_S\ge 0.71$, which corresponds to $\sim 4$dB of squeezing at the detector side.

{\em Correcting the displacement noise.---}While CV-tele-QT protocol does not require any encoding, we can further combine it with quantum error correction to protect the quantum information against additive noise. We consider the case of $\kappa_H=\kappa_S=1$ for simplicity, while the general case can be solved similarly. Given Eq.~\eqref{add_channel_G}, the additive noise is asymmetric between the two quadratures. Consequently, the suitable code to consider is a rectangle lattice Gottesman-Kitaev-Preskill (GKP) code~\cite{gottesman2001,brady2024advances}, with lattice spacing $L_q,L_p$ for position and momentum quadratures satisfying the ratio $L_q/L_p=\sqrt{(1-\eta)/\eta}$. For example, consider a single qubit GKP code, under the convention that $\hbar=2$, the corresponding spacings $L_q=(\frac{1-\eta}{\eta})^{1/4}2\sqrt{2\pi}$ and $L_p=(\frac{1-\eta}{\eta})^{-1/4}2\sqrt{2\pi}$, leading to the logical states $\ket{b}_L=\sum_{n\in \mathbb{Z}}\ket{(2n+b)L_q/2}_q$, where the summation is over all integers $n$ and $\ket{x}_q$ is the (infinitely-squeezed) position eigenstate with eigenvalue $x$. Displacement errors smaller than half of the grid size can be corrected, therefore we can correct displacement errors within the grid of $[-L_q/2,L_q/2]\times[-L_p/2,L_p/2]$, leading to the error probability
$
P_E\leq  1-{\rm Erf}[\sqrt{G\pi}(\eta(1-\eta))^{1/4}]^2\sim \exp(-\sqrt{\eta(1-\eta)}G\pi),
$ where the last step of asymptotic assumes $G\gg1$.

{\em Discussions.---}The proposed generalized CV teleportation protocol can also be adopted for entanglement swap, via introducing an additional ancilla $I$ entangled with $S$. Given that the quantum state of $S$ can be teleported through to the output $B^\prime$, the entanglement between $S$ and $I$ can then be transformed to entanglement between $I$ and $B^\prime$, completing an entanglement swap process from $AB$, $SI$ to $BI$. Although generalized CV teleportation in other forms have been proposed in prior works~\cite{horoshko2000quantum,mivsta2005improving} motivated by non-demolition interaction, none of them considered quantum transduction.

The proposed CV-tele-QT protocol provides enhancement in quantum transduction, while only requiring offline squeezing, homodyne measurement and displacement. The enhancement is substantial especially in the low-efficiency region. While at a higher efficiency, it is an open question whether the proposed protocol can be improved to maintain the enhancement.

The proposed protocol differs from previous proposals of enhancing quantum transduction. Ref.~\cite{wu2021deterministic} adopts traditional CV teleportation, while relying on alternation of the quantum transduction device to generate microwave-optical entanglement. Our proposal does not require microwave-optical entanglement to be generated. Previous adaptive protocol~\cite{zhang2018quantum} and intraband entanglement-assisted protocol~\cite{shi2024overcoming} requires inline squeezing, which is challenging in experiments, while our approach only relies on offline squeezing. Furthermore, the current protocol only requires $\eta>(1-\sqrt{1-e^2/G^2})/2$, which $\sim 1/G^2$ when $G$ is large, to have a positive rate (viz. Eq.~\eqref{G_star}), while Ref.~\cite{shi2024overcoming} requires $\eta>1/(G+1)$.

\begin{acknowledgements}
QZ acknowledges discussions with Pengcheng Liao and Zheshen Zhang. This work is supported by Office of Naval Research Grant No. N00014-23-1-2296, National Science Foundation OMA-2326746, National Science Foundation CAREER Award CCF-2240641, DARPA D24AC00153-02 and an unrestricted gift from Google. 
\end{acknowledgements}

{\em Appendix: proof of Theorem 1---} 
The Wigner function of the joint state $\hat{\rho}_{S^\prime A^\prime B}$ right before the homodyne measurement and corrective displacement
\begin{align} 
&W(q_{S^\prime},p_{S^\prime},q_{A^\prime},p_{A^\prime},{\bm x}_B;\hat{\rho}_{S^\prime A^\prime B})=\nonumber
\\
&\ \  W(\sqrt{1-\eta}q_{S^\prime}-\sqrt{\eta}q_{A^\prime},\sqrt{1-\eta}p_{S^\prime}-\sqrt{\eta}p_{A^\prime};\hat{\rho}_{S})\times
\nonumber
\\
&\ \  W(\sqrt{\eta}q_{S^\prime}+\sqrt{1-\eta}q_{A^\prime},\sqrt{\eta}p_{S^\prime}+\sqrt{1-\eta}p_{A^\prime},{\bm x}_B;\hat\psi_{AB}),
\label{Wigner_all}
\end{align}
where we denote ${\bm x}_B=(q_B,p_B)$ and $\hat{\rho}_S$ is the (unknown) quantum state of the signal $S$ that the sender aims to transfer towards the receiver.
This is obtained via considering the beamsplitter transform in Eq.~\eqref{AS_beamsplitter}.
Conditioned on the measurement results $\tilde{q}$ and $\tilde{p}$, the Wigner function of the output state after the corrective displacement is
$
W({\bm x}_B|\tilde{q},\tilde{p}; \hat{\rho}_{B^\prime})
\propto \int dq_{S^\prime}dp_{A^\prime}  W(q_{S^\prime},\tilde{p},\tilde{q},p_{A^\prime},{\bm x}_B+\tilde{\bm d}_B;\hat{\rho}_{S^\prime A^\prime B})$, 
where we have denoted the corrective displacement as ${\tilde{\bm d}}_B=(\tilde{q}/\sqrt{\eta},-\tilde{p}/\sqrt{1-\eta})$. Throughout the proof, the integration interval is $[-\infty,\infty]$ for all variables. Utilizing Eq.~\eqref{Wigner_all}, we have
\begin{align} 
&W({\bm x}_B|\tilde{q},\tilde{p}; \hat{\rho}_{B^\prime})\nonumber\\
&\propto \int dq_{S^\prime}dp_{A^\prime}W(\sqrt{1-\eta}q_{S^\prime}-\sqrt{\eta}\tilde{q},\sqrt{1-\eta}\tilde{p}-\sqrt{\eta}p_{A^\prime};\hat{\rho}_{S})
\nonumber
\\
&\times W(\sqrt{\eta}q_{S^\prime}+\sqrt{1-\eta}\tilde{q},\sqrt{\eta}\tilde{p}+\sqrt{1-\eta}p_{A^\prime},{\bm x}_B+\tilde{\bm d}_B;\hat\psi_{AB}).
\end{align}
With a change of variable, $q_S=\sqrt{1-\eta}q_{S^\prime}-\sqrt{\eta}\tilde{q}$ and $p_S=\sqrt{1-\eta}\tilde{p}-\sqrt{\eta}p_{A^\prime}$, we have a simpler form
\begin{align}
&W({\bm x}_B|\tilde{q},\tilde{p}; \hat{\rho}_{B^\prime})\propto \int dq_{S}dp_{S}W(q_S,p_S;\hat{\rho}_{S})
\nonumber
\\
&\times W(\frac{\sqrt{\eta}q_S+\tilde{q}}{\sqrt{1-\eta}},\frac{\tilde{p}-\sqrt{1-\eta}p_S}{\sqrt{\eta}},{\bm x}_B+\tilde{\bm d}_B;\hat\psi_{AB}).
\label{W_cond}
\end{align}

In the infinite squeezing limit of $G\to \infty$, the Wigner function 
$W(q_A,p_A,q_B,p_B;\hat{\psi}_{AB})
\propto\delta(\sqrt{1-\eta}q_A-\sqrt{\eta}q_B)\delta(\sqrt{\eta}p_A+\sqrt{1-\eta}p_B)$. 
Inputting the above into Eq.~\eqref{W_cond}, this immediately leads to
$
W({\bm x}_B|\tilde{q},\tilde{p}; \hat{\rho}_{B^\prime})=W(q_B,p_B;\hat{\rho}_S),
$
and we have proven Eq.~\eqref{equality_ginf} of Theorem~\ref{theorem:cv_teleportation}.


Now we proceed to the finite squeezing case.
From the Wigner function of single-mode squeezed vacuum of gain $G$ and Eq.~\eqref{generalized_EPR}, we have 
\begin{align} 
&W(q_A,p_A,{\bm x}_B;\hat{\psi}_{AB})\propto \nonumber
\\
&\exp\left[{-\frac{(\sqrt{1-\eta}q_A-\sqrt{\eta}q_B)^2+(\sqrt{\eta}p_A+\sqrt{1-\eta}p_B)^2}{2/G}}\right]\nonumber
\\
&
\times \exp\left[{-\frac{(\sqrt{1-\eta}p_A-\sqrt{\eta}p_B)^2+(\sqrt{\eta}q_A+\sqrt{1-\eta}q_B)^2}{2G}}\right].
\label{Wig_finiteG}
\end{align} 
For simplicity, we directly integrate over the different measurement outcomes to obtain the Wigner function of the unconditional state
$
W({\bm x}_B; \hat{\rho}_{B^\prime})
=\int d\tilde{q}d\tilde{p} dq_{S^\prime}dp_{A^\prime}  W(q_{S^\prime},\tilde{p},\tilde{q},p_{A^\prime},{\bm x}_B+\tilde{\bm d}_B;\hat{\rho}_{S^\prime A^\prime B}).
$
Adopting the finite-squeezing Wigner function of Eq.~\eqref{Wig_finiteG} into Eq.~\eqref{Wigner_all} and performing the same change-of-variable in Eq.~\eqref{W_cond}, we have
\begin{widetext}
\begin{align}
&W(q_B,p_B; \hat{\rho}_{B^\prime})\propto  \int d\tilde{q}d\tilde{p}\int dq_{S}dp_{S}W(q_S,p_S;\hat{\rho}_{S})\nonumber
\\
&\quad \times \exp [-\frac{\eta\left(q_S-q_B\right)^2+\left(1-\eta\right)\left(p_B-p_S\right)^2}{2/G}-\frac{\left(\frac{\tilde{p}}{\sqrt{\eta(1-\eta)}}-\frac{(1-\eta )p_S+\eta p_B}{\sqrt{\eta}}\right)^2+\left(\frac{\tilde{q}}{\sqrt{(1-\eta)\eta}}+\frac{\eta q_S+(1-\eta)q_B}{\sqrt{1-\eta}}\right)^2}{2G}]
\\
&\propto \int dq_{S}dp_{S}W(q_S,p_S;\hat{\rho}_{S})\exp\left[-\frac{\left(q_S-q_B\right)^2}{2/\eta G}-\frac{\left(p_B-p_S\right)^2}{2/\left(1-\eta\right)G}\right].
\end{align}
\end{widetext}
This is an asymmetric additive noise channel with variances $1/\eta G$ and $1/(1-\eta)G$. Therefore, we conclude our proof for Theorem~\ref{fig:schematic}.

Now we extend to the case with experimental imperfections. The homodyne loss $1-\kappa_H$ can be moved before the transduction device as we assume it is identical on $S^\prime$ and $A^\prime$. Therefore, now we are equivalently teleportating the lossy input $\calL_{\kappa_H}(\hat{\rho}_S)$ with the lossy entangled states $\calL_{\kappa_H \kappa_S}\otimes \calL_{\kappa_S} (\hat{\psi}_{AB})$. Eq.~\eqref{W_cond} can therefore be updated as
\begin{align}
&W({\bm x}_B|\tilde{q},\tilde{p}; \hat{\rho}_{B^\prime})\propto \int dq_{S}dp_{S}W(q_S,p_S;\calL_{\kappa_H}(\hat{\rho}_S))
\nonumber
\\
&\times W(\frac{\sqrt{\eta}q_S+\tilde{q}}{\sqrt{1-\eta}},\frac{\tilde{p}-\sqrt{1-\eta}p_S}{\sqrt{\eta}},{\bm x}_B+\tilde{\bm d}_B^{\rm L};\calL_{\kappa_H \kappa_S}\otimes \calL_{\kappa_S} (\hat{\psi}_{AB})).
\label{W_cond_loss}
\end{align}
The corrective displacement ${\tilde{\bm d}}_B^L=(\tilde{q}/\sqrt{\kappa_H\eta},-\tilde{p}/\sqrt{\kappa_H(1-\eta)})$ is chosen accordingly to amplify the signal for countering the loss. Similar to the perfect case, we integrate out the measurement results $\tilde{q}$ and $\tilde{p}$ and obtain the output Wigner function
\begin{widetext}
\begin{align}
W(q_B,p_B; \hat{\rho}_{B^\prime})&\propto \int dq_{S}dp_{S}
W(q_S,p_S;\calL_{\kappa_H}(\hat{\rho}_{S}))\nonumber
\\
&\times
\exp\left[-\frac{\left(q_S-\sqrt{\kappa_H}q_B\right)^2\eta}{2\left[1-\eta+\kappa_H\left(\eta+\left(1/G-1\right)\kappa_S\right)\right]}-\frac{\left(p_S-\sqrt{\kappa_H}p_B\right)^2(1-\eta)}{2\left[\kappa_H+(1/G-1)\kappa_H\kappa_S+\eta(1-\kappa_H)\right]}\right]
\\
&\propto \int dq_{S}dp_{S}\int d^2{\bm x}
W(\bm x;\hat{\rho}_{S})\exp\left[-\frac{(\bm x-{\bm x}_S/\sqrt{\kappa_H})^2}{2(1-\kappa_H)/\kappa_H}\right]\nonumber
\\
&\times
\exp\left[-\frac{\left(q_S-\sqrt{\kappa_H}q_B\right)^2\eta}{2\left(1-\eta+\kappa_H\left(\eta+\left(1/G-1\right)\kappa_S\right)\right)}-\frac{\left(p_S-\sqrt{\kappa_H}p_B\right)^2(1-\eta)}{2(\kappa_H+(1/G-1)\kappa_H\kappa_S+\eta(1-\kappa_H))}\right]
\\
&\propto \int dq dp
\exp\left[
-\frac{\kappa_H\eta (q-q_B)^2}{2\left[\kappa_H\kappa_S/G+\left(1-\kappa_H\kappa_S\right)\right]}
-
\frac{\kappa_H(1-\eta)(p-p_B)^2}{2\left[\kappa_H\kappa_S/G+\left(1-\kappa_H\kappa_S\right)\right]}\right].
\end{align}
The corresponding additive noises are
\begin{align}
&\sigma_q^2=\left[\kappa_H\kappa_S/G+\left(1-\kappa_H\kappa_S\right)\right]/(\eta\kappa_H),
\\
&\sigma_p^2=\left[\kappa_H\kappa_S/G+\left(1-\kappa_H\kappa_S\right)\right]/((1-\eta)\kappa_H).
\end{align}

\end{widetext}


\begin{thebibliography}{43}%
\makeatletter
\providecommand \@ifxundefined [1]{%
 \@ifx{#1\undefined}
}%
\providecommand \@ifnum [1]{%
 \ifnum #1\expandafter \@firstoftwo
 \else \expandafter \@secondoftwo
 \fi
}%
\providecommand \@ifx [1]{%
 \ifx #1\expandafter \@firstoftwo
 \else \expandafter \@secondoftwo
 \fi
}%
\providecommand \natexlab [1]{#1}%
\providecommand \enquote  [1]{``#1''}%
\providecommand \bibnamefont  [1]{#1}%
\providecommand \bibfnamefont [1]{#1}%
\providecommand \citenamefont [1]{#1}%
\providecommand \href@noop [0]{\@secondoftwo}%
\providecommand \href [0]{\begingroup \@sanitize@url \@href}%
\providecommand \@href[1]{\@@startlink{#1}\@@href}%
\providecommand \@@href[1]{\endgroup#1\@@endlink}%
\providecommand \@sanitize@url [0]{\catcode `\\12\catcode `\$12\catcode
  `\&12\catcode `\#12\catcode `\^12\catcode `\_12\catcode `\%12\relax}%
\providecommand \@@startlink[1]{}%
\providecommand \@@endlink[0]{}%
\providecommand \url  [0]{\begingroup\@sanitize@url \@url }%
\providecommand \@url [1]{\endgroup\@href {#1}{\urlprefix }}%
\providecommand \urlprefix  [0]{URL }%
\providecommand \Eprint [0]{\href }%
\providecommand \doibase [0]{https://doi.org/}%
\providecommand \selectlanguage [0]{\@gobble}%
\providecommand \bibinfo  [0]{\@secondoftwo}%
\providecommand \bibfield  [0]{\@secondoftwo}%
\providecommand \translation [1]{[#1]}%
\providecommand \BibitemOpen [0]{}%
\providecommand \bibitemStop [0]{}%
\providecommand \bibitemNoStop [0]{.\EOS\space}%
\providecommand \EOS [0]{\spacefactor3000\relax}%
\providecommand \BibitemShut  [1]{\csname bibitem#1\endcsname}%
\let\auto@bib@innerbib\@empty
\bibitem [{\citenamefont {Lauk}\ \emph {et~al.}(2020)\citenamefont {Lauk},
  \citenamefont {Sinclair}, \citenamefont {Barzanjeh}, \citenamefont {Covey},
  \citenamefont {Saffman}, \citenamefont {Spiropulu},\ and\ \citenamefont
  {Simon}}]{lauk2020perspectives}%
  \BibitemOpen
  \bibfield  {author} {\bibinfo {author} {\bibfnamefont {N.}~\bibnamefont
  {Lauk}}, \bibinfo {author} {\bibfnamefont {N.}~\bibnamefont {Sinclair}},
  \bibinfo {author} {\bibfnamefont {S.}~\bibnamefont {Barzanjeh}}, \bibinfo
  {author} {\bibfnamefont {J.~P.}\ \bibnamefont {Covey}}, \bibinfo {author}
  {\bibfnamefont {M.}~\bibnamefont {Saffman}}, \bibinfo {author} {\bibfnamefont
  {M.}~\bibnamefont {Spiropulu}},\ and\ \bibinfo {author} {\bibfnamefont
  {C.}~\bibnamefont {Simon}},\ }\bibfield  {title} {\bibinfo {title}
  {Perspectives on quantum transduction},\ }\href@noop {} {\bibfield  {journal}
  {\bibinfo  {journal} {Quantum Sci. Techno.}\ }\textbf {\bibinfo {volume}
  {5}},\ \bibinfo {pages} {020501} (\bibinfo {year} {2020})}\BibitemShut
  {NoStop}%
\bibitem [{\citenamefont {Awschalom}\ \emph {et~al.}(2021)\citenamefont
  {Awschalom}, \citenamefont {Berggren}, \citenamefont {Bernien}, \citenamefont
  {Bhave}, \citenamefont {Carr}, \citenamefont {Davids}, \citenamefont
  {Economou}, \citenamefont {Englund}, \citenamefont {Faraon}, \citenamefont
  {Fejer} \emph {et~al.}}]{awschalom2021development}%
  \BibitemOpen
  \bibfield  {author} {\bibinfo {author} {\bibfnamefont {D.}~\bibnamefont
  {Awschalom}}, \bibinfo {author} {\bibfnamefont {K.~K.}\ \bibnamefont
  {Berggren}}, \bibinfo {author} {\bibfnamefont {H.}~\bibnamefont {Bernien}},
  \bibinfo {author} {\bibfnamefont {S.}~\bibnamefont {Bhave}}, \bibinfo
  {author} {\bibfnamefont {L.~D.}\ \bibnamefont {Carr}}, \bibinfo {author}
  {\bibfnamefont {P.}~\bibnamefont {Davids}}, \bibinfo {author} {\bibfnamefont
  {S.~E.}\ \bibnamefont {Economou}}, \bibinfo {author} {\bibfnamefont
  {D.}~\bibnamefont {Englund}}, \bibinfo {author} {\bibfnamefont
  {A.}~\bibnamefont {Faraon}}, \bibinfo {author} {\bibfnamefont
  {M.}~\bibnamefont {Fejer}}, \emph {et~al.},\ }\bibfield  {title} {\bibinfo
  {title} {Development of quantum interconnects (quics) for next-generation
  information technologies},\ }\href@noop {} {\bibfield  {journal} {\bibinfo
  {journal} {PRX Quantum}\ }\textbf {\bibinfo {volume} {2}},\ \bibinfo {pages}
  {017002} (\bibinfo {year} {2021})}\BibitemShut {NoStop}%
\bibitem [{\citenamefont {Han}\ \emph {et~al.}(2021)\citenamefont {Han},
  \citenamefont {Fu}, \citenamefont {Zou}, \citenamefont {Jiang},\ and\
  \citenamefont {Tang}}]{han2021microwave}%
  \BibitemOpen
  \bibfield  {author} {\bibinfo {author} {\bibfnamefont {X.}~\bibnamefont
  {Han}}, \bibinfo {author} {\bibfnamefont {W.}~\bibnamefont {Fu}}, \bibinfo
  {author} {\bibfnamefont {C.-L.}\ \bibnamefont {Zou}}, \bibinfo {author}
  {\bibfnamefont {L.}~\bibnamefont {Jiang}},\ and\ \bibinfo {author}
  {\bibfnamefont {H.~X.}\ \bibnamefont {Tang}},\ }\bibfield  {title} {\bibinfo
  {title} {Microwave-optical quantum frequency conversion},\ }\href@noop {}
  {\bibfield  {journal} {\bibinfo  {journal} {Optica}\ }\textbf {\bibinfo
  {volume} {8}},\ \bibinfo {pages} {1050} (\bibinfo {year} {2021})}\BibitemShut
  {NoStop}%
\bibitem [{\citenamefont {Andrews}\ \emph {et~al.}(2014)\citenamefont
  {Andrews}, \citenamefont {Peterson}, \citenamefont {Purdy}, \citenamefont
  {Cicak}, \citenamefont {Simmonds}, \citenamefont {Regal},\ and\ \citenamefont
  {Lehnert}}]{andrews2014bidirectional}%
  \BibitemOpen
  \bibfield  {author} {\bibinfo {author} {\bibfnamefont {R.~W.}\ \bibnamefont
  {Andrews}}, \bibinfo {author} {\bibfnamefont {R.~W.}\ \bibnamefont
  {Peterson}}, \bibinfo {author} {\bibfnamefont {T.~P.}\ \bibnamefont {Purdy}},
  \bibinfo {author} {\bibfnamefont {K.}~\bibnamefont {Cicak}}, \bibinfo
  {author} {\bibfnamefont {R.~W.}\ \bibnamefont {Simmonds}}, \bibinfo {author}
  {\bibfnamefont {C.~A.}\ \bibnamefont {Regal}},\ and\ \bibinfo {author}
  {\bibfnamefont {K.~W.}\ \bibnamefont {Lehnert}},\ }\bibfield  {title}
  {\bibinfo {title} {Bidirectional and efficient conversion between microwave
  and optical light},\ }\href@noop {} {\bibfield  {journal} {\bibinfo
  {journal} {Nat. Phys.}\ }\textbf {\bibinfo {volume} {10}},\ \bibinfo {pages}
  {321} (\bibinfo {year} {2014})}\BibitemShut {NoStop}%
\bibitem [{\citenamefont {Bochmann}\ \emph {et~al.}(2013)\citenamefont
  {Bochmann}, \citenamefont {Vainsencher}, \citenamefont {Awschalom},\ and\
  \citenamefont {Cleland}}]{bochmann2013nanomechanical}%
  \BibitemOpen
  \bibfield  {author} {\bibinfo {author} {\bibfnamefont {J.}~\bibnamefont
  {Bochmann}}, \bibinfo {author} {\bibfnamefont {A.}~\bibnamefont
  {Vainsencher}}, \bibinfo {author} {\bibfnamefont {D.~D.}\ \bibnamefont
  {Awschalom}},\ and\ \bibinfo {author} {\bibfnamefont {A.~N.}\ \bibnamefont
  {Cleland}},\ }\bibfield  {title} {\bibinfo {title} {Nanomechanical coupling
  between microwave and optical photons},\ }\href@noop {} {\bibfield  {journal}
  {\bibinfo  {journal} {Nat. Phys.}\ }\textbf {\bibinfo {volume} {9}},\
  \bibinfo {pages} {712} (\bibinfo {year} {2013})}\BibitemShut {NoStop}%
\bibitem [{\citenamefont {Vainsencher}\ \emph {et~al.}(2016)\citenamefont
  {Vainsencher}, \citenamefont {Satzinger}, \citenamefont {Peairs},\ and\
  \citenamefont {Cleland}}]{vainsencher2016bi}%
  \BibitemOpen
  \bibfield  {author} {\bibinfo {author} {\bibfnamefont {A.}~\bibnamefont
  {Vainsencher}}, \bibinfo {author} {\bibfnamefont {K.}~\bibnamefont
  {Satzinger}}, \bibinfo {author} {\bibfnamefont {G.}~\bibnamefont {Peairs}},\
  and\ \bibinfo {author} {\bibfnamefont {A.}~\bibnamefont {Cleland}},\
  }\bibfield  {title} {\bibinfo {title} {Bi-directional conversion between
  microwave and optical frequencies in a piezoelectric optomechanical device},\
  }\href@noop {} {\bibfield  {journal} {\bibinfo  {journal} {Appl. Phys.
  Lett.}\ }\textbf {\bibinfo {volume} {109}},\ \bibinfo {pages} {033107}
  (\bibinfo {year} {2016})}\BibitemShut {NoStop}%
\bibitem [{\citenamefont {Balram}\ \emph {et~al.}(2016)\citenamefont {Balram},
  \citenamefont {Davan{\c{c}}o}, \citenamefont {Song},\ and\ \citenamefont
  {Srinivasan}}]{balram2016coherent}%
  \BibitemOpen
  \bibfield  {author} {\bibinfo {author} {\bibfnamefont {K.~C.}\ \bibnamefont
  {Balram}}, \bibinfo {author} {\bibfnamefont {M.~I.}\ \bibnamefont
  {Davan{\c{c}}o}}, \bibinfo {author} {\bibfnamefont {J.~D.}\ \bibnamefont
  {Song}},\ and\ \bibinfo {author} {\bibfnamefont {K.}~\bibnamefont
  {Srinivasan}},\ }\bibfield  {title} {\bibinfo {title} {Coherent coupling
  between radiofrequency, optical and acoustic waves in piezo-optomechanical
  circuits},\ }\href@noop {} {\bibfield  {journal} {\bibinfo  {journal} {Nat.
  Photonics}\ }\textbf {\bibinfo {volume} {10}},\ \bibinfo {pages} {346}
  (\bibinfo {year} {2016})}\BibitemShut {NoStop}%
\bibitem [{\citenamefont {Tsang}(2010)}]{Tsang2010}%
  \BibitemOpen
  \bibfield  {author} {\bibinfo {author} {\bibfnamefont {M.}~\bibnamefont
  {Tsang}},\ }\bibfield  {title} {\bibinfo {title} {Cavity quantum
  electro-optics},\ }\href {https://doi.org/10.1103/PhysRevA.81.063837}
  {\bibfield  {journal} {\bibinfo  {journal} {Phys. Rev. A}\ }\textbf {\bibinfo
  {volume} {81}},\ \bibinfo {pages} {063837} (\bibinfo {year}
  {2010})}\BibitemShut {NoStop}%
\bibitem [{\citenamefont {Tsang}(2011)}]{Tsang2011}%
  \BibitemOpen
  \bibfield  {author} {\bibinfo {author} {\bibfnamefont {M.}~\bibnamefont
  {Tsang}},\ }\bibfield  {title} {\bibinfo {title} {Cavity quantum
  electro-optics. ii. input-output relations between traveling optical and
  microwave fields},\ }\href {https://doi.org/10.1103/PhysRevA.84.043845}
  {\bibfield  {journal} {\bibinfo  {journal} {Phys. Rev. A}\ }\textbf {\bibinfo
  {volume} {84}},\ \bibinfo {pages} {043845} (\bibinfo {year}
  {2011})}\BibitemShut {NoStop}%
\bibitem [{\citenamefont {Fan}\ \emph {et~al.}(2018)\citenamefont {Fan},
  \citenamefont {Zou}, \citenamefont {Cheng}, \citenamefont {Guo},
  \citenamefont {Han}, \citenamefont {Gong}, \citenamefont {Wang},\ and\
  \citenamefont {Tang}}]{fan2018superconducting}%
  \BibitemOpen
  \bibfield  {author} {\bibinfo {author} {\bibfnamefont {L.}~\bibnamefont
  {Fan}}, \bibinfo {author} {\bibfnamefont {C.-L.}\ \bibnamefont {Zou}},
  \bibinfo {author} {\bibfnamefont {R.}~\bibnamefont {Cheng}}, \bibinfo
  {author} {\bibfnamefont {X.}~\bibnamefont {Guo}}, \bibinfo {author}
  {\bibfnamefont {X.}~\bibnamefont {Han}}, \bibinfo {author} {\bibfnamefont
  {Z.}~\bibnamefont {Gong}}, \bibinfo {author} {\bibfnamefont {S.}~\bibnamefont
  {Wang}},\ and\ \bibinfo {author} {\bibfnamefont {H.~X.}\ \bibnamefont
  {Tang}},\ }\bibfield  {title} {\bibinfo {title} {Superconducting cavity
  electro-optics: a platform for coherent photon conversion between
  superconducting and photonic circuits},\ }\href@noop {} {\bibfield  {journal}
  {\bibinfo  {journal} {Sci. Adv.}\ }\textbf {\bibinfo {volume} {4}},\ \bibinfo
  {pages} {eaar4994} (\bibinfo {year} {2018})}\BibitemShut {NoStop}%
\bibitem [{\citenamefont {Xu}\ \emph {et~al.}(2020)\citenamefont {Xu},
  \citenamefont {Sayem}, \citenamefont {Fan}, \citenamefont {Wang},
  \citenamefont {Cheng}, \citenamefont {Zou}, \citenamefont {Fu}, \citenamefont
  {Yang}, \citenamefont {Xu},\ and\ \citenamefont
  {Tang}}]{xu2020bidirectional}%
  \BibitemOpen
  \bibfield  {author} {\bibinfo {author} {\bibfnamefont {Y.}~\bibnamefont
  {Xu}}, \bibinfo {author} {\bibfnamefont {A.~A.}\ \bibnamefont {Sayem}},
  \bibinfo {author} {\bibfnamefont {L.}~\bibnamefont {Fan}}, \bibinfo {author}
  {\bibfnamefont {S.}~\bibnamefont {Wang}}, \bibinfo {author} {\bibfnamefont
  {R.}~\bibnamefont {Cheng}}, \bibinfo {author} {\bibfnamefont {C.-L.}\
  \bibnamefont {Zou}}, \bibinfo {author} {\bibfnamefont {W.}~\bibnamefont
  {Fu}}, \bibinfo {author} {\bibfnamefont {L.}~\bibnamefont {Yang}}, \bibinfo
  {author} {\bibfnamefont {M.}~\bibnamefont {Xu}},\ and\ \bibinfo {author}
  {\bibfnamefont {H.~X.}\ \bibnamefont {Tang}},\ }\bibfield  {title} {\bibinfo
  {title} {Bidirectional electro-optic conversion reaching 1\% efficiency with
  thin-film lithium niobate},\ }\href@noop {} {\bibfield  {journal} {\bibinfo
  {journal} {arXiv:2012.14909}\ } (\bibinfo {year} {2020})}\BibitemShut
  {NoStop}%
\bibitem [{\citenamefont {Jiang}\ \emph {et~al.}(2020)\citenamefont {Jiang},
  \citenamefont {Sarabalis}, \citenamefont {Dahmani}, \citenamefont {Patel},
  \citenamefont {Mayor}, \citenamefont {McKenna}, \citenamefont {Van~Laer},\
  and\ \citenamefont {Safavi-Naeini}}]{jiang2020efficient}%
  \BibitemOpen
  \bibfield  {author} {\bibinfo {author} {\bibfnamefont {W.}~\bibnamefont
  {Jiang}}, \bibinfo {author} {\bibfnamefont {C.~J.}\ \bibnamefont
  {Sarabalis}}, \bibinfo {author} {\bibfnamefont {Y.~D.}\ \bibnamefont
  {Dahmani}}, \bibinfo {author} {\bibfnamefont {R.~N.}\ \bibnamefont {Patel}},
  \bibinfo {author} {\bibfnamefont {F.~M.}\ \bibnamefont {Mayor}}, \bibinfo
  {author} {\bibfnamefont {T.~P.}\ \bibnamefont {McKenna}}, \bibinfo {author}
  {\bibfnamefont {R.}~\bibnamefont {Van~Laer}},\ and\ \bibinfo {author}
  {\bibfnamefont {A.~H.}\ \bibnamefont {Safavi-Naeini}},\ }\bibfield  {title}
  {\bibinfo {title} {Efficient bidirectional piezo-optomechanical transduction
  between microwave and optical frequency},\ }\href@noop {} {\bibfield
  {journal} {\bibinfo  {journal} {Nat. Commun.}\ }\textbf {\bibinfo {volume}
  {11}},\ \bibinfo {pages} {1} (\bibinfo {year} {2020})}\BibitemShut {NoStop}%
\bibitem [{\citenamefont {Verd\'u}\ \emph {et~al.}(2009)\citenamefont
  {Verd\'u}, \citenamefont {Zoubi}, \citenamefont {Koller}, \citenamefont
  {Majer}, \citenamefont {Ritsch},\ and\ \citenamefont
  {Schmiedmayer}}]{PhysRevLett.103.043603}%
  \BibitemOpen
  \bibfield  {author} {\bibinfo {author} {\bibfnamefont {J.}~\bibnamefont
  {Verd\'u}}, \bibinfo {author} {\bibfnamefont {H.}~\bibnamefont {Zoubi}},
  \bibinfo {author} {\bibfnamefont {C.}~\bibnamefont {Koller}}, \bibinfo
  {author} {\bibfnamefont {J.}~\bibnamefont {Majer}}, \bibinfo {author}
  {\bibfnamefont {H.}~\bibnamefont {Ritsch}},\ and\ \bibinfo {author}
  {\bibfnamefont {J.}~\bibnamefont {Schmiedmayer}},\ }\bibfield  {title}
  {\bibinfo {title} {Strong magnetic coupling of an ultracold gas to a
  superconducting waveguide cavity},\ }\href
  {https://doi.org/10.1103/PhysRevLett.103.043603} {\bibfield  {journal}
  {\bibinfo  {journal} {Phys. Rev. Lett.}\ }\textbf {\bibinfo {volume} {103}},\
  \bibinfo {pages} {043603} (\bibinfo {year} {2009})}\BibitemShut {NoStop}%
\bibitem [{\citenamefont {Williamson}\ \emph {et~al.}(2014)\citenamefont
  {Williamson}, \citenamefont {Chen},\ and\ \citenamefont
  {Longdell}}]{PhysRevLett.113.203601}%
  \BibitemOpen
  \bibfield  {author} {\bibinfo {author} {\bibfnamefont {L.~A.}\ \bibnamefont
  {Williamson}}, \bibinfo {author} {\bibfnamefont {Y.-H.}\ \bibnamefont
  {Chen}},\ and\ \bibinfo {author} {\bibfnamefont {J.~J.}\ \bibnamefont
  {Longdell}},\ }\bibfield  {title} {\bibinfo {title} {Magneto-optic modulator
  with unit quantum efficiency},\ }\href
  {https://doi.org/10.1103/PhysRevLett.113.203601} {\bibfield  {journal}
  {\bibinfo  {journal} {Phys. Rev. Lett.}\ }\textbf {\bibinfo {volume} {113}},\
  \bibinfo {pages} {203601} (\bibinfo {year} {2014})}\BibitemShut {NoStop}%
\bibitem [{\citenamefont {Shao}\ \emph {et~al.}(2019)\citenamefont {Shao},
  \citenamefont {Yu}, \citenamefont {Maity}, \citenamefont {Sinclair},
  \citenamefont {Zheng}, \citenamefont {Chia}, \citenamefont {Shams-Ansari},
  \citenamefont {Wang}, \citenamefont {Zhang}, \citenamefont {Lai} \emph
  {et~al.}}]{shao2019microwave}%
  \BibitemOpen
  \bibfield  {author} {\bibinfo {author} {\bibfnamefont {L.}~\bibnamefont
  {Shao}}, \bibinfo {author} {\bibfnamefont {M.}~\bibnamefont {Yu}}, \bibinfo
  {author} {\bibfnamefont {S.}~\bibnamefont {Maity}}, \bibinfo {author}
  {\bibfnamefont {N.}~\bibnamefont {Sinclair}}, \bibinfo {author}
  {\bibfnamefont {L.}~\bibnamefont {Zheng}}, \bibinfo {author} {\bibfnamefont
  {C.}~\bibnamefont {Chia}}, \bibinfo {author} {\bibfnamefont {A.}~\bibnamefont
  {Shams-Ansari}}, \bibinfo {author} {\bibfnamefont {C.}~\bibnamefont {Wang}},
  \bibinfo {author} {\bibfnamefont {M.}~\bibnamefont {Zhang}}, \bibinfo
  {author} {\bibfnamefont {K.}~\bibnamefont {Lai}}, \emph {et~al.},\ }\bibfield
   {title} {\bibinfo {title} {Microwave-to-optical conversion using lithium
  niobate thin-film acoustic resonators},\ }\href@noop {} {\bibfield  {journal}
  {\bibinfo  {journal} {Optica}\ }\textbf {\bibinfo {volume} {6}},\ \bibinfo
  {pages} {1498} (\bibinfo {year} {2019})}\BibitemShut {NoStop}%
\bibitem [{\citenamefont {Fiaschi}\ \emph {et~al.}(2021)\citenamefont
  {Fiaschi}, \citenamefont {Hensen}, \citenamefont {Wallucks}, \citenamefont
  {Benevides}, \citenamefont {Li}, \citenamefont {Alegre},\ and\ \citenamefont
  {Gr{\"o}blacher}}]{fiaschi2021optomechanical}%
  \BibitemOpen
  \bibfield  {author} {\bibinfo {author} {\bibfnamefont {N.}~\bibnamefont
  {Fiaschi}}, \bibinfo {author} {\bibfnamefont {B.}~\bibnamefont {Hensen}},
  \bibinfo {author} {\bibfnamefont {A.}~\bibnamefont {Wallucks}}, \bibinfo
  {author} {\bibfnamefont {R.}~\bibnamefont {Benevides}}, \bibinfo {author}
  {\bibfnamefont {J.}~\bibnamefont {Li}}, \bibinfo {author} {\bibfnamefont
  {T.~P.~M.}\ \bibnamefont {Alegre}},\ and\ \bibinfo {author} {\bibfnamefont
  {S.}~\bibnamefont {Gr{\"o}blacher}},\ }\bibfield  {title} {\bibinfo {title}
  {Optomechanical quantum teleportation},\ }\href@noop {} {\bibfield  {journal}
  {\bibinfo  {journal} {Nat. Photon.}\ }\textbf {\bibinfo {volume} {15}},\
  \bibinfo {pages} {817} (\bibinfo {year} {2021})}\BibitemShut {NoStop}%
\bibitem [{\citenamefont {Han}\ \emph {et~al.}(2020)\citenamefont {Han},
  \citenamefont {Fu}, \citenamefont {Zhong}, \citenamefont {Zou}, \citenamefont
  {Xu}, \citenamefont {Al~Sayem}, \citenamefont {Xu}, \citenamefont {Wang},
  \citenamefont {Cheng}, \citenamefont {Jiang} \emph {et~al.}}]{han2020cavity}%
  \BibitemOpen
  \bibfield  {author} {\bibinfo {author} {\bibfnamefont {X.}~\bibnamefont
  {Han}}, \bibinfo {author} {\bibfnamefont {W.}~\bibnamefont {Fu}}, \bibinfo
  {author} {\bibfnamefont {C.}~\bibnamefont {Zhong}}, \bibinfo {author}
  {\bibfnamefont {C.-L.}\ \bibnamefont {Zou}}, \bibinfo {author} {\bibfnamefont
  {Y.}~\bibnamefont {Xu}}, \bibinfo {author} {\bibfnamefont {A.}~\bibnamefont
  {Al~Sayem}}, \bibinfo {author} {\bibfnamefont {M.}~\bibnamefont {Xu}},
  \bibinfo {author} {\bibfnamefont {S.}~\bibnamefont {Wang}}, \bibinfo {author}
  {\bibfnamefont {R.}~\bibnamefont {Cheng}}, \bibinfo {author} {\bibfnamefont
  {L.}~\bibnamefont {Jiang}}, \emph {et~al.},\ }\bibfield  {title} {\bibinfo
  {title} {Cavity piezo-mechanics for superconducting-nanophotonic quantum
  interface},\ }\href@noop {} {\bibfield  {journal} {\bibinfo  {journal} {Nat.
  Commun.}\ }\textbf {\bibinfo {volume} {11}},\ \bibinfo {pages} {1} (\bibinfo
  {year} {2020})}\BibitemShut {NoStop}%
\bibitem [{\citenamefont {Zhong}\ \emph {et~al.}(2020)\citenamefont {Zhong},
  \citenamefont {Wang}, \citenamefont {Zou}, \citenamefont {Zhang},
  \citenamefont {Han}, \citenamefont {Fu}, \citenamefont {Xu}, \citenamefont
  {Shankar}, \citenamefont {Devoret}, \citenamefont {Tang} \emph
  {et~al.}}]{zhong2020proposal}%
  \BibitemOpen
  \bibfield  {author} {\bibinfo {author} {\bibfnamefont {C.}~\bibnamefont
  {Zhong}}, \bibinfo {author} {\bibfnamefont {Z.}~\bibnamefont {Wang}},
  \bibinfo {author} {\bibfnamefont {C.}~\bibnamefont {Zou}}, \bibinfo {author}
  {\bibfnamefont {M.}~\bibnamefont {Zhang}}, \bibinfo {author} {\bibfnamefont
  {X.}~\bibnamefont {Han}}, \bibinfo {author} {\bibfnamefont {W.}~\bibnamefont
  {Fu}}, \bibinfo {author} {\bibfnamefont {M.}~\bibnamefont {Xu}}, \bibinfo
  {author} {\bibfnamefont {S.}~\bibnamefont {Shankar}}, \bibinfo {author}
  {\bibfnamefont {M.~H.}\ \bibnamefont {Devoret}}, \bibinfo {author}
  {\bibfnamefont {H.~X.}\ \bibnamefont {Tang}}, \emph {et~al.},\ }\bibfield
  {title} {\bibinfo {title} {Proposal for heralded generation and detection of
  entangled microwave--optical-photon pairs},\ }\href@noop {} {\bibfield
  {journal} {\bibinfo  {journal} {Phys. Rev. Lett.}\ }\textbf {\bibinfo
  {volume} {124}},\ \bibinfo {pages} {010511} (\bibinfo {year}
  {2020})}\BibitemShut {NoStop}%
\bibitem [{\citenamefont {Mirhosseini}\ \emph {et~al.}(2020)\citenamefont
  {Mirhosseini}, \citenamefont {Sipahigil}, \citenamefont {Kalaee},\ and\
  \citenamefont {Painter}}]{mirhosseini2020superconducting}%
  \BibitemOpen
  \bibfield  {author} {\bibinfo {author} {\bibfnamefont {M.}~\bibnamefont
  {Mirhosseini}}, \bibinfo {author} {\bibfnamefont {A.}~\bibnamefont
  {Sipahigil}}, \bibinfo {author} {\bibfnamefont {M.}~\bibnamefont {Kalaee}},\
  and\ \bibinfo {author} {\bibfnamefont {O.}~\bibnamefont {Painter}},\
  }\bibfield  {title} {\bibinfo {title} {Superconducting qubit to optical
  photon transduction},\ }\href@noop {} {\bibfield  {journal} {\bibinfo
  {journal} {Nature}\ }\textbf {\bibinfo {volume} {588}},\ \bibinfo {pages}
  {599} (\bibinfo {year} {2020})}\BibitemShut {NoStop}%
\bibitem [{\citenamefont {Forsch}\ \emph {et~al.}(2020)\citenamefont {Forsch},
  \citenamefont {Stockill}, \citenamefont {Wallucks}, \citenamefont
  {Marinkovi{\'c}}, \citenamefont {G{\"a}rtner}, \citenamefont {Norte},
  \citenamefont {van Otten}, \citenamefont {Fiore}, \citenamefont
  {Srinivasan},\ and\ \citenamefont {Gr{\"o}blacher}}]{forsch2020microwave}%
  \BibitemOpen
  \bibfield  {author} {\bibinfo {author} {\bibfnamefont {M.}~\bibnamefont
  {Forsch}}, \bibinfo {author} {\bibfnamefont {R.}~\bibnamefont {Stockill}},
  \bibinfo {author} {\bibfnamefont {A.}~\bibnamefont {Wallucks}}, \bibinfo
  {author} {\bibfnamefont {I.}~\bibnamefont {Marinkovi{\'c}}}, \bibinfo
  {author} {\bibfnamefont {C.}~\bibnamefont {G{\"a}rtner}}, \bibinfo {author}
  {\bibfnamefont {R.~A.}\ \bibnamefont {Norte}}, \bibinfo {author}
  {\bibfnamefont {F.}~\bibnamefont {van Otten}}, \bibinfo {author}
  {\bibfnamefont {A.}~\bibnamefont {Fiore}}, \bibinfo {author} {\bibfnamefont
  {K.}~\bibnamefont {Srinivasan}},\ and\ \bibinfo {author} {\bibfnamefont
  {S.}~\bibnamefont {Gr{\"o}blacher}},\ }\bibfield  {title} {\bibinfo {title}
  {Microwave-to-optics conversion using a mechanical oscillator in its quantum
  ground state},\ }\href@noop {} {\bibfield  {journal} {\bibinfo  {journal}
  {Nat. Phys.}\ }\textbf {\bibinfo {volume} {16}},\ \bibinfo {pages} {69}
  (\bibinfo {year} {2020})}\BibitemShut {NoStop}%
\bibitem [{\citenamefont {Holzgrafe}\ \emph {et~al.}(2020)\citenamefont
  {Holzgrafe}, \citenamefont {Sinclair}, \citenamefont {Zhu}, \citenamefont
  {Shams-Ansari}, \citenamefont {Colangelo}, \citenamefont {Hu}, \citenamefont
  {Zhang}, \citenamefont {Berggren},\ and\ \citenamefont
  {Lon{\v{c}}ar}}]{holzgrafe2020cavity}%
  \BibitemOpen
  \bibfield  {author} {\bibinfo {author} {\bibfnamefont {J.}~\bibnamefont
  {Holzgrafe}}, \bibinfo {author} {\bibfnamefont {N.}~\bibnamefont {Sinclair}},
  \bibinfo {author} {\bibfnamefont {D.}~\bibnamefont {Zhu}}, \bibinfo {author}
  {\bibfnamefont {A.}~\bibnamefont {Shams-Ansari}}, \bibinfo {author}
  {\bibfnamefont {M.}~\bibnamefont {Colangelo}}, \bibinfo {author}
  {\bibfnamefont {Y.}~\bibnamefont {Hu}}, \bibinfo {author} {\bibfnamefont
  {M.}~\bibnamefont {Zhang}}, \bibinfo {author} {\bibfnamefont {K.~K.}\
  \bibnamefont {Berggren}},\ and\ \bibinfo {author} {\bibfnamefont
  {M.}~\bibnamefont {Lon{\v{c}}ar}},\ }\bibfield  {title} {\bibinfo {title}
  {Cavity electro-optics in thin-film lithium niobate for efficient
  microwave-to-optical transduction},\ }\href@noop {} {\bibfield  {journal}
  {\bibinfo  {journal} {Optica}\ }\textbf {\bibinfo {volume} {7}},\ \bibinfo
  {pages} {1714} (\bibinfo {year} {2020})}\BibitemShut {NoStop}%
\bibitem [{\citenamefont {Sahu}\ \emph {et~al.}(2022)\citenamefont {Sahu},
  \citenamefont {Hease}, \citenamefont {Rueda}, \citenamefont {Arnold},
  \citenamefont {Qiu},\ and\ \citenamefont {Fink}}]{sahu2022quantum}%
  \BibitemOpen
  \bibfield  {author} {\bibinfo {author} {\bibfnamefont {R.}~\bibnamefont
  {Sahu}}, \bibinfo {author} {\bibfnamefont {W.}~\bibnamefont {Hease}},
  \bibinfo {author} {\bibfnamefont {A.}~\bibnamefont {Rueda}}, \bibinfo
  {author} {\bibfnamefont {G.}~\bibnamefont {Arnold}}, \bibinfo {author}
  {\bibfnamefont {L.}~\bibnamefont {Qiu}},\ and\ \bibinfo {author}
  {\bibfnamefont {J.~M.}\ \bibnamefont {Fink}},\ }\bibfield  {title} {\bibinfo
  {title} {Quantum-enabled operation of a microwave-optical interface},\
  }\href@noop {} {\bibfield  {journal} {\bibinfo  {journal} {Nat. Commun.}\
  }\textbf {\bibinfo {volume} {13}},\ \bibinfo {pages} {1276} (\bibinfo {year}
  {2022})}\BibitemShut {NoStop}%
\bibitem [{\citenamefont {Brubaker}\ \emph {et~al.}(2022)\citenamefont
  {Brubaker}, \citenamefont {Kindem}, \citenamefont {Urmey}, \citenamefont
  {Mittal}, \citenamefont {Delaney}, \citenamefont {Burns}, \citenamefont
  {Vissers}, \citenamefont {Lehnert},\ and\ \citenamefont
  {Regal}}]{brubaker2022optomechanical}%
  \BibitemOpen
  \bibfield  {author} {\bibinfo {author} {\bibfnamefont {B.~M.}\ \bibnamefont
  {Brubaker}}, \bibinfo {author} {\bibfnamefont {J.~M.}\ \bibnamefont
  {Kindem}}, \bibinfo {author} {\bibfnamefont {M.~D.}\ \bibnamefont {Urmey}},
  \bibinfo {author} {\bibfnamefont {S.}~\bibnamefont {Mittal}}, \bibinfo
  {author} {\bibfnamefont {R.~D.}\ \bibnamefont {Delaney}}, \bibinfo {author}
  {\bibfnamefont {P.~S.}\ \bibnamefont {Burns}}, \bibinfo {author}
  {\bibfnamefont {M.~R.}\ \bibnamefont {Vissers}}, \bibinfo {author}
  {\bibfnamefont {K.~W.}\ \bibnamefont {Lehnert}},\ and\ \bibinfo {author}
  {\bibfnamefont {C.~A.}\ \bibnamefont {Regal}},\ }\bibfield  {title} {\bibinfo
  {title} {Optomechanical ground-state cooling in a continuous and efficient
  electro-optic transducer},\ }\href@noop {} {\bibfield  {journal} {\bibinfo
  {journal} {Phys. Rev. X}\ }\textbf {\bibinfo {volume} {12}},\ \bibinfo
  {pages} {021062} (\bibinfo {year} {2022})}\BibitemShut {NoStop}%
\bibitem [{\citenamefont {Qiu}\ \emph {et~al.}(2023)\citenamefont {Qiu},
  \citenamefont {Sahu}, \citenamefont {Hease}, \citenamefont {Arnold},\ and\
  \citenamefont {Fink}}]{qiu2023coherent}%
  \BibitemOpen
  \bibfield  {author} {\bibinfo {author} {\bibfnamefont {L.}~\bibnamefont
  {Qiu}}, \bibinfo {author} {\bibfnamefont {R.}~\bibnamefont {Sahu}}, \bibinfo
  {author} {\bibfnamefont {W.}~\bibnamefont {Hease}}, \bibinfo {author}
  {\bibfnamefont {G.}~\bibnamefont {Arnold}},\ and\ \bibinfo {author}
  {\bibfnamefont {J.~M.}\ \bibnamefont {Fink}},\ }\bibfield  {title} {\bibinfo
  {title} {Coherent optical control of a superconducting microwave cavity via
  electro-optical dynamical back-action},\ }\href@noop {} {\bibfield  {journal}
  {\bibinfo  {journal} {Nat. Commun.}\ }\textbf {\bibinfo {volume} {14}},\
  \bibinfo {pages} {3784} (\bibinfo {year} {2023})}\BibitemShut {NoStop}%
\bibitem [{\citenamefont {Sahu}\ \emph {et~al.}(2023)\citenamefont {Sahu},
  \citenamefont {Qiu}, \citenamefont {Hease}, \citenamefont {Arnold},
  \citenamefont {Minoguchi}, \citenamefont {Rabl},\ and\ \citenamefont
  {Fink}}]{sahu2023entangling}%
  \BibitemOpen
  \bibfield  {author} {\bibinfo {author} {\bibfnamefont {R.}~\bibnamefont
  {Sahu}}, \bibinfo {author} {\bibfnamefont {L.}~\bibnamefont {Qiu}}, \bibinfo
  {author} {\bibfnamefont {W.}~\bibnamefont {Hease}}, \bibinfo {author}
  {\bibfnamefont {G.}~\bibnamefont {Arnold}}, \bibinfo {author} {\bibfnamefont
  {Y.}~\bibnamefont {Minoguchi}}, \bibinfo {author} {\bibfnamefont
  {P.}~\bibnamefont {Rabl}},\ and\ \bibinfo {author} {\bibfnamefont {J.~M.}\
  \bibnamefont {Fink}},\ }\bibfield  {title} {\bibinfo {title} {Entangling
  microwaves with light},\ }\href@noop {} {\bibfield  {journal} {\bibinfo
  {journal} {Science}\ }\textbf {\bibinfo {volume} {380}},\ \bibinfo {pages}
  {718} (\bibinfo {year} {2023})}\BibitemShut {NoStop}%
\bibitem [{\citenamefont {Bennett}\ \emph {et~al.}(1993)\citenamefont
  {Bennett}, \citenamefont {Brassard}, \citenamefont {Cr{\'e}peau},
  \citenamefont {Jozsa}, \citenamefont {Peres},\ and\ \citenamefont
  {Wootters}}]{bennett1993}%
  \BibitemOpen
  \bibfield  {author} {\bibinfo {author} {\bibfnamefont {C.~H.}\ \bibnamefont
  {Bennett}}, \bibinfo {author} {\bibfnamefont {G.}~\bibnamefont {Brassard}},
  \bibinfo {author} {\bibfnamefont {C.}~\bibnamefont {Cr{\'e}peau}}, \bibinfo
  {author} {\bibfnamefont {R.}~\bibnamefont {Jozsa}}, \bibinfo {author}
  {\bibfnamefont {A.}~\bibnamefont {Peres}},\ and\ \bibinfo {author}
  {\bibfnamefont {W.~K.}\ \bibnamefont {Wootters}},\ }\bibfield  {title}
  {\bibinfo {title} {Teleporting an unknown quantum state via dual classical
  and einstein-podolsky-rosen channels},\ }\href@noop {} {\bibfield  {journal}
  {\bibinfo  {journal} {Phys. Rev. Lett.}\ }\textbf {\bibinfo {volume} {70}},\
  \bibinfo {pages} {1895} (\bibinfo {year} {1993})}\BibitemShut {NoStop}%
\bibitem [{\citenamefont {Braunstein}\ and\ \citenamefont
  {Kimble}(1998)}]{braunstein1998}%
  \BibitemOpen
  \bibfield  {author} {\bibinfo {author} {\bibfnamefont {S.~L.}\ \bibnamefont
  {Braunstein}}\ and\ \bibinfo {author} {\bibfnamefont {H.~J.}\ \bibnamefont
  {Kimble}},\ }\bibfield  {title} {\bibinfo {title} {Teleportation of
  continuous quantum variables},\ }\href@noop {} {\bibfield  {journal}
  {\bibinfo  {journal} {Phys. Rev. Lett.}\ }\textbf {\bibinfo {volume} {80}},\
  \bibinfo {pages} {869} (\bibinfo {year} {1998})}\BibitemShut {NoStop}%
\bibitem [{\citenamefont {Pirandola}\ and\ \citenamefont
  {Mancini}(2006)}]{pirandola2006}%
  \BibitemOpen
  \bibfield  {author} {\bibinfo {author} {\bibfnamefont {S.}~\bibnamefont
  {Pirandola}}\ and\ \bibinfo {author} {\bibfnamefont {S.}~\bibnamefont
  {Mancini}},\ }\bibfield  {title} {\bibinfo {title} {Quantum teleportation
  with continuous variables: A survey},\ }\href@noop {} {\bibfield  {journal}
  {\bibinfo  {journal} {Laser Phys.}\ }\textbf {\bibinfo {volume} {16}},\
  \bibinfo {pages} {1418} (\bibinfo {year} {2006})}\BibitemShut {NoStop}%
\bibitem [{\citenamefont {Furusawa}\ \emph {et~al.}(1998)\citenamefont
  {Furusawa}, \citenamefont {S{\o}rensen}, \citenamefont {Braunstein},
  \citenamefont {Fuchs}, \citenamefont {Kimble},\ and\ \citenamefont
  {Polzik}}]{furusawa1998unconditional}%
  \BibitemOpen
  \bibfield  {author} {\bibinfo {author} {\bibfnamefont {A.}~\bibnamefont
  {Furusawa}}, \bibinfo {author} {\bibfnamefont {J.~L.}\ \bibnamefont
  {S{\o}rensen}}, \bibinfo {author} {\bibfnamefont {S.~L.}\ \bibnamefont
  {Braunstein}}, \bibinfo {author} {\bibfnamefont {C.~A.}\ \bibnamefont
  {Fuchs}}, \bibinfo {author} {\bibfnamefont {H.~J.}\ \bibnamefont {Kimble}},\
  and\ \bibinfo {author} {\bibfnamefont {E.~S.}\ \bibnamefont {Polzik}},\
  }\bibfield  {title} {\bibinfo {title} {Unconditional quantum teleportation},\
  }\href@noop {} {\bibfield  {journal} {\bibinfo  {journal} {Science}\ }\textbf
  {\bibinfo {volume} {282}},\ \bibinfo {pages} {706} (\bibinfo {year}
  {1998})}\BibitemShut {NoStop}%
\bibitem [{\citenamefont {Wu}\ \emph {et~al.}(2021)\citenamefont {Wu},
  \citenamefont {Cui}, \citenamefont {Fan},\ and\ \citenamefont
  {Zhuang}}]{wu2021deterministic}%
  \BibitemOpen
  \bibfield  {author} {\bibinfo {author} {\bibfnamefont {J.}~\bibnamefont
  {Wu}}, \bibinfo {author} {\bibfnamefont {C.}~\bibnamefont {Cui}}, \bibinfo
  {author} {\bibfnamefont {L.}~\bibnamefont {Fan}},\ and\ \bibinfo {author}
  {\bibfnamefont {Q.}~\bibnamefont {Zhuang}},\ }\bibfield  {title} {\bibinfo
  {title} {Deterministic microwave-optical transduction based on quantum
  teleportation},\ }\href {https://doi.org/10.1103/PhysRevApplied.16.064044}
  {\bibfield  {journal} {\bibinfo  {journal} {Phys. Rev. Appl.}\ }\textbf
  {\bibinfo {volume} {16}},\ \bibinfo {pages} {064044} (\bibinfo {year}
  {2021})}\BibitemShut {NoStop}%
\bibitem [{\citenamefont {Weedbrook}\ \emph {et~al.}(2012)\citenamefont
  {Weedbrook}, \citenamefont {Pirandola}, \citenamefont {Garc\'{\i}a-Patr\'on},
  \citenamefont {Cerf}, \citenamefont {Ralph}, \citenamefont {Shapiro},\ and\
  \citenamefont {Lloyd}}]{Weedbrook2012}%
  \BibitemOpen
  \bibfield  {author} {\bibinfo {author} {\bibfnamefont {C.}~\bibnamefont
  {Weedbrook}}, \bibinfo {author} {\bibfnamefont {S.}~\bibnamefont
  {Pirandola}}, \bibinfo {author} {\bibfnamefont {R.}~\bibnamefont
  {Garc\'{\i}a-Patr\'on}}, \bibinfo {author} {\bibfnamefont {N.~J.}\
  \bibnamefont {Cerf}}, \bibinfo {author} {\bibfnamefont {T.~C.}\ \bibnamefont
  {Ralph}}, \bibinfo {author} {\bibfnamefont {J.~H.}\ \bibnamefont {Shapiro}},\
  and\ \bibinfo {author} {\bibfnamefont {S.}~\bibnamefont {Lloyd}},\ }\bibfield
   {title} {\bibinfo {title} {Gaussian quantum information},\ }\href
  {https://doi.org/10.1103/RevModPhys.84.621} {\bibfield  {journal} {\bibinfo
  {journal} {Rev. Mod. Phys.}\ }\textbf {\bibinfo {volume} {84}},\ \bibinfo
  {pages} {621} (\bibinfo {year} {2012})}\BibitemShut {NoStop}%
\bibitem [{\citenamefont {Lloyd}(1997)}]{lloyd1997capacity}%
  \BibitemOpen
  \bibfield  {author} {\bibinfo {author} {\bibfnamefont {S.}~\bibnamefont
  {Lloyd}},\ }\bibfield  {title} {\bibinfo {title} {Capacity of the noisy
  quantum channel},\ }\href@noop {} {\bibfield  {journal} {\bibinfo  {journal}
  {Phys. Rev. A}\ }\textbf {\bibinfo {volume} {55}},\ \bibinfo {pages} {1613}
  (\bibinfo {year} {1997})}\BibitemShut {NoStop}%
\bibitem [{\citenamefont {Shor}(2002)}]{shor2002quantum}%
  \BibitemOpen
  \bibfield  {author} {\bibinfo {author} {\bibfnamefont {P.~W.}\ \bibnamefont
  {Shor}},\ }\bibfield  {title} {\bibinfo {title} {The quantum channel capacity
  and coherent information},\ }in\ \href@noop {} {\emph {\bibinfo {booktitle}
  {lecture notes, MSRI Workshop on Quantum Computation}}}\ (\bibinfo {year}
  {2002})\BibitemShut {NoStop}%
\bibitem [{\citenamefont {Devetak}(2005)}]{devetak2005private}%
  \BibitemOpen
  \bibfield  {author} {\bibinfo {author} {\bibfnamefont {I.}~\bibnamefont
  {Devetak}},\ }\bibfield  {title} {\bibinfo {title} {The private classical
  capacity and quantum capacity of a quantum channel},\ }\href@noop {}
  {\bibfield  {journal} {\bibinfo  {journal} {IEEE Trans. Inf. Theory}\
  }\textbf {\bibinfo {volume} {51}},\ \bibinfo {pages} {44} (\bibinfo {year}
  {2005})}\BibitemShut {NoStop}%
\bibitem [{\citenamefont {Holevo}\ and\ \citenamefont
  {Werner}(2001)}]{holevo2001evaluating}%
  \BibitemOpen
  \bibfield  {author} {\bibinfo {author} {\bibfnamefont {A.~S.}\ \bibnamefont
  {Holevo}}\ and\ \bibinfo {author} {\bibfnamefont {R.~F.}\ \bibnamefont
  {Werner}},\ }\bibfield  {title} {\bibinfo {title} {Evaluating capacities of
  bosonic gaussian channels},\ }\href@noop {} {\bibfield  {journal} {\bibinfo
  {journal} {Phys. Rev. A}\ }\textbf {\bibinfo {volume} {63}},\ \bibinfo
  {pages} {032312} (\bibinfo {year} {2001})}\BibitemShut {NoStop}%
\bibitem [{\citenamefont {Wu}\ and\ \citenamefont
  {Zhuang}(2021)}]{wu2021continuous}%
  \BibitemOpen
  \bibfield  {author} {\bibinfo {author} {\bibfnamefont {J.}~\bibnamefont
  {Wu}}\ and\ \bibinfo {author} {\bibfnamefont {Q.}~\bibnamefont {Zhuang}},\
  }\bibfield  {title} {\bibinfo {title} {Continuous-variable error correction
  for general gaussian noises},\ }\href@noop {} {\bibfield  {journal} {\bibinfo
   {journal} {Phys. Rev. Appl.}\ }\textbf {\bibinfo {volume} {15}},\ \bibinfo
  {pages} {034073} (\bibinfo {year} {2021})}\BibitemShut {NoStop}%
\bibitem [{\citenamefont {Fanizza}\ \emph {et~al.}(2021)\citenamefont
  {Fanizza}, \citenamefont {Kianvash},\ and\ \citenamefont
  {Giovannetti}}]{fanizza2021estimating}%
  \BibitemOpen
  \bibfield  {author} {\bibinfo {author} {\bibfnamefont {M.}~\bibnamefont
  {Fanizza}}, \bibinfo {author} {\bibfnamefont {F.}~\bibnamefont {Kianvash}},\
  and\ \bibinfo {author} {\bibfnamefont {V.}~\bibnamefont {Giovannetti}},\
  }\bibfield  {title} {\bibinfo {title} {Estimating quantum and private
  capacities of gaussian channels via degradable extensions},\ }\href@noop {}
  {\bibfield  {journal} {\bibinfo  {journal} {arXiv:2103.09569}\ } (\bibinfo
  {year} {2021})}\BibitemShut {NoStop}%
\bibitem [{\citenamefont {Gottesman}\ \emph {et~al.}(2001)\citenamefont
  {Gottesman}, \citenamefont {Kitaev},\ and\ \citenamefont
  {Preskill}}]{gottesman2001}%
  \BibitemOpen
  \bibfield  {author} {\bibinfo {author} {\bibfnamefont {D.}~\bibnamefont
  {Gottesman}}, \bibinfo {author} {\bibfnamefont {A.}~\bibnamefont {Kitaev}},\
  and\ \bibinfo {author} {\bibfnamefont {J.}~\bibnamefont {Preskill}},\
  }\bibfield  {title} {\bibinfo {title} {Encoding a qubit in an oscillator},\
  }\href {https://doi.org/10.1103/PhysRevA.64.012310} {\bibfield  {journal}
  {\bibinfo  {journal} {Phys. Rev. A}\ }\textbf {\bibinfo {volume} {64}},\
  \bibinfo {pages} {012310} (\bibinfo {year} {2001})}\BibitemShut {NoStop}%
\bibitem [{\citenamefont {Brady}\ \emph {et~al.}(2024)\citenamefont {Brady},
  \citenamefont {Eickbusch}, \citenamefont {Singh}, \citenamefont {Wu},\ and\
  \citenamefont {Zhuang}}]{brady2024advances}%
  \BibitemOpen
  \bibfield  {author} {\bibinfo {author} {\bibfnamefont {A.~J.}\ \bibnamefont
  {Brady}}, \bibinfo {author} {\bibfnamefont {A.}~\bibnamefont {Eickbusch}},
  \bibinfo {author} {\bibfnamefont {S.}~\bibnamefont {Singh}}, \bibinfo
  {author} {\bibfnamefont {J.}~\bibnamefont {Wu}},\ and\ \bibinfo {author}
  {\bibfnamefont {Q.}~\bibnamefont {Zhuang}},\ }\bibfield  {title} {\bibinfo
  {title} {Advances in bosonic quantum error correction with
  gottesman--kitaev--preskill codes: Theory, engineering and applications},\
  }\href@noop {} {\bibfield  {journal} {\bibinfo  {journal} {Progress in
  Quantum Electronics}\ }\textbf {\bibinfo {volume} {93}},\ \bibinfo {pages}
  {100496} (\bibinfo {year} {2024})}\BibitemShut {NoStop}%
\bibitem [{\citenamefont {Horoshko}\ and\ \citenamefont
  {Kilin}(2000)}]{horoshko2000quantum}%
  \BibitemOpen
  \bibfield  {author} {\bibinfo {author} {\bibfnamefont {D.}~\bibnamefont
  {Horoshko}}\ and\ \bibinfo {author} {\bibfnamefont {S.~Y.}\ \bibnamefont
  {Kilin}},\ }\bibfield  {title} {\bibinfo {title} {Quantum teleportation using
  quantum nondemolition technique},\ }\href@noop {} {\bibfield  {journal}
  {\bibinfo  {journal} {Phys. Rev. A}\ }\textbf {\bibinfo {volume} {61}},\
  \bibinfo {pages} {032304} (\bibinfo {year} {2000})}\BibitemShut {NoStop}%
\bibitem [{\citenamefont {Mi{\v{s}}ta~Jr}\ and\ \citenamefont
  {Filip}(2005)}]{mivsta2005improving}%
  \BibitemOpen
  \bibfield  {author} {\bibinfo {author} {\bibfnamefont {L.}~\bibnamefont
  {Mi{\v{s}}ta~Jr}}\ and\ \bibinfo {author} {\bibfnamefont {R.}~\bibnamefont
  {Filip}},\ }\bibfield  {title} {\bibinfo {title} {Improving teleportation of
  continuous variables by local operations},\ }\href@noop {} {\bibfield
  {journal} {\bibinfo  {journal} {Phys. Rev. A}\ }\textbf {\bibinfo {volume}
  {71}},\ \bibinfo {pages} {032342} (\bibinfo {year} {2005})}\BibitemShut
  {NoStop}%
\bibitem [{\citenamefont {Zhang}\ \emph {et~al.}(2018)\citenamefont {Zhang},
  \citenamefont {Zou},\ and\ \citenamefont {Jiang}}]{zhang2018quantum}%
  \BibitemOpen
  \bibfield  {author} {\bibinfo {author} {\bibfnamefont {M.}~\bibnamefont
  {Zhang}}, \bibinfo {author} {\bibfnamefont {C.-L.}\ \bibnamefont {Zou}},\
  and\ \bibinfo {author} {\bibfnamefont {L.}~\bibnamefont {Jiang}},\ }\bibfield
   {title} {\bibinfo {title} {Quantum transduction with adaptive control},\
  }\href@noop {} {\bibfield  {journal} {\bibinfo  {journal} {Phys. Rev. Lett.}\
  }\textbf {\bibinfo {volume} {120}},\ \bibinfo {pages} {020502} (\bibinfo
  {year} {2018})}\BibitemShut {NoStop}%
\bibitem [{\citenamefont {Shi}\ and\ \citenamefont
  {Zhuang}(2024)}]{shi2024overcoming}%
  \BibitemOpen
  \bibfield  {author} {\bibinfo {author} {\bibfnamefont {H.}~\bibnamefont
  {Shi}}\ and\ \bibinfo {author} {\bibfnamefont {Q.}~\bibnamefont {Zhuang}},\
  }\bibfield  {title} {\bibinfo {title} {Overcoming the fundamental limit of
  quantum transduction via intraband entanglement},\ }\href@noop {} {\bibfield
  {journal} {\bibinfo  {journal} {Optica Quantum}\ }\textbf {\bibinfo {volume}
  {2}},\ \bibinfo {pages} {475} (\bibinfo {year} {2024})}\BibitemShut {NoStop}%
\end{thebibliography}
\end{document}